\documentclass[12pt]{article}
\usepackage{amsmath,amssymb, graphicx, booktabs, hyperref, fancyhdr, titlesec, float, listings, xcolor, amsthm, array, arydshln}
\usepackage{enumitem}
\usepackage{algorithm}
\usepackage{algpseudocode}
\usepackage{adjustbox}
\usepackage{tablefootnote}
\usepackage{multirow}

\usepackage[margin=1in]{geometry}
\newtheorem{theorem}{Theorem}[section]

\newtheorem{definition}[theorem]{Definition}

\definecolor{codegreen}{rgb}{0,0.6,0}
\definecolor{codegray}{rgb}{0.5,0.5,0.5}
\definecolor{codepurple}{rgb}{0.58,0,0.82}
\definecolor{backcolour}{rgb}{0.95,0.95,0.92}

\hypersetup{
    colorlinks=true,        
    linkcolor=blue,         
    citecolor=blue,          
    urlcolor=blue           
}

\lstdefinestyle{mystyle}{
    backgroundcolor=\color{backcolour},   
    commentstyle=\color{codegreen},
    keywordstyle=\color{magenta},
    numberstyle=\tiny\color{codegray},
    stringstyle=\color{codepurple},
    basicstyle=\ttfamily\small,
    breakatwhitespace=false,         
    breaklines=true,                 
    captionpos=b,                    
    keepspaces=true,                 
    numbers=left,                    
    numbersep=5pt,                  
    showspaces=false,                
    showstringspaces=false,
    showtabs=false,                  
    tabsize=2
}

\title{Applications of higher order Markov models and Pressure Index to strategize controlled run chases in Twenty20 cricket}
\author{Rhitankar Bandyopadhyay \footnote{Email: \href{mailto:rhitankar.isi@gmail.com}{rhitankar.isi@gmail.com}} 
\\Indian Statistical Institute, Kolkata, India \\ 
\\ Dibyojyoti Bhattacharjee\footnote{Email: \href{mailto:djb.stat@gmail.com}{djb.stat@gmail.com}}
\\Department of Statistics, Assam University, Silchar, India
}
\date{}

\begin{document}

\maketitle

\begin{abstract}
In limited overs cricket, the team batting first posts a target score for the team batting second to achieve in order to win the match. The team batting second is constrained by decreasing resources in terms of number of balls left and number of wickets in hand in the process of reaching the target as the second innings progresses. The Pressure Index, a measure created by researchers in the past, serves as a tool for quantifying the level of pressure that a team batting second encounters in limited overs cricket. Through a ball-by-ball analysis of the second innings, it reveals how effectively the team batting second in a limited-over game proceeds towards their target. This research employs higher order Markov chains to examine the strategies employed by successful teams during run chases in Twenty20 matches. By studying the trends in successful run chases spanning over $16$ years and utilizing a significant dataset of $6537$ Twenty20 matches, specific strategies are identified. Consequently, an efficient approach to successful run chases in Twenty20 cricket is formulated, effectively limiting the Pressure Index to $[0.5, 3.5]$ or even further down under $0.5$ as early as possible. The innovative methodology adopted in this research offers valuable insights for cricket teams looking to enhance their performance in run chases.

\textbf{Keywords}: Cricket analytics, Data mining in sports, Markov chain, Markov model, Pressure Index, Transition matrix 
\end{abstract}

\newpage

\section{Introduction}\label{sec:introduction}
Cricket is a team sport played between two teams consisting of 11 players each, following a set of rules as per the format of the game being Tests, One Day Internationals or Twenty20 Internationals (First Class, List A and Twenty20 respectively on a larger basis). In addition, shorter formats of the game like T10, The Hundred, etc., are being introduced for rapid spread and popularity of the game worldwide. One of the major reasons for the intense popularity of cricket nowadays is the uncertainty during a game and it has been observed that the shorter the format, the higher the uncertainty. The above claim is consistent with the fact that in the $5$-year period from July $2014$ to June $2019$, $29.89\ \%$ Test matches have been won by teams outside the top $3$ ICC men's Test rankings against teams ranked within the top $3$ ICC men's Test rankings, which increases to $33.67\ \%$ if we consider Twenty20 cricket during the same period. The rise is even more in the following $5$-year period from July $2019$ to June $2024$ where it increases from $19.48\ \%$ in Tests to $29.30\ \%$ in Twenty20 cricket\footnote{The data for the results of these matches have been collected from \url{https://stats.espncricinfo.com/ci/engine/stats/index.html}}.

In Twenty20 cricket, a team bats for a maximum of 20 overs with 10 wickets in hand and tries to score as many runs as possible within the provided resources. The team which fields first then comes in to bat with identical resources, being set a target to score more runs than the team which batted first. Unless in rain-interrupted matches or other scenarios resulting in reduction of overs, the team which scores more runs wins the match. The match is called a tie in case both teams end up scoring equal number of runs. Super overs (earlier, bowl outs) are often played to decide the outcome of tied matches.

As the game progresses, efficient use of numbers, statistics and technology plays a key role to analyze and strategize the game from a player and team’s point of view. The present study is an endeavor to introduce a strategy whereby teams could constantly set a target, within themselves, for their batters at certain intervals throughout the innings while chasing a score in a regular Twenty20 match with the help of the Pressure Index and higher order Markov chains. The Pressure Index is a ball-by-ball measure which quantifies the pressure experienced by the team batting second. It is a function of the required run rate and wickets lost at any point in the second innings, for our purposes, in a Twenty20 match. Thus, during a run chase, the value of the Pressure Index might change after every ball. The value of the Pressure Index starts at unity at the start of a run chase and keeps on changing as the innings progresses. It eventually reaches zero if the target is achieved but rises steeply if the team batting second fails to achieve the target. Details on working formulae, functionality, etc., regarding the Pressure Index are discussed later. The curve obtained by plotting the Pressure Index values against each ball during an innings is said to be the Pressure Curve. The Pressure Curve gives a clear indication of the status of the batting team at any given instance during a run chase. In this paper, Pressure Index values have been obtained corresponding to numerous Twenty20 matches spanned over a prolonged period. Then the concepts of Markov chains have been used in a way to learn the pattern of changes in the Pressure Index throughout a run chase in Twenty20 cricket and an attempt has been made to suggest a suitable strategy to keep the Pressure Index under control so that teams do not often lose grip on run chases.

\subsection{Basic terminologies}\label{sec:terminologies}

A few basic terms prevalant in Twenty20 cricket are described as follows.

\begin{enumerate}
\renewcommand{\labelenumi}{\roman{enumi}.}
    \item \textit{Overs}: In cricket, the batsman stands at one end of the pitch, and the bowler delivers the ball from the other end. Now, after the bowler has delivered six consecutive legal balls, he is said to have completed an over.

    \item \textit{Powerplay}: The first $6$ overs of a Twenty20 match are called powerplay, and a maximum of two fielders can be placed outside the 30-yard circle during this period (see \cite{talukdar2020investigating}).

    \item \textit{Death Overs}: The last few overs of any innings of a cricket match are said to be Death Overs. We will typically denote the last $4$ overs of a Twenty20 innings to be Death Overs throughout the paper.

    \item \textit{Middle Overs}: The overs following the powerplay till the beginning of the death overs is said to be Middle Overs.

    \item \textit{Openers}: Since batting takes place in pairs in cricket, the pair of batters who come in to bat at the start of an innings are termed as \textit{Opening partners} or simply, openers (see \cite{talukdar2020investigating}).

    \item \textit{Batting Depth}: If the bowlers of a team, who generally come to bat lower down the order, also possess good batting skills, then the team is said to have significant batting depth.

    \item \textit{Tailenders}: Players who are participating in the team primarily for bowling skills and aren't a master in the art of batting are called Tailenders or \textit{late-order batters} (see \cite{knight2023cricket}).

    \item \textit{Target}: In limited overs cricket, the team batting first (refer as Team $A$) tries to score as many runs as they can in the allocated number of overs, without losing all their wickets. At the end of the innings, one run more than the total number of runs scored by Team $A$ is the target for the opposition (Team $B$, say). Team $B$ has to reach the target runs within the alloted number of overs in order to win the match.
\end{enumerate}

 Following a literature review and discussions on the novelty of this work in Section~\ref{sec:literature}, in Section~\ref{sec:markovchain}, we present the fundamental concepts of Markov chains and motivate the need for higher-order models in the context of run-chase dynamics. Section~\ref{sec:pi} introduces the Pressure Index in detail, outlines its computation, and demonstrates how its over-wise evolution reflects match situations. In Section~\ref{sec:estimating-order}, we estimate the optimal order of the proposed Markov model. In Section~\ref{sec:discretization}, we discuss discretization of Pressure Index suggesting suitable gridding. Sections~\ref{sec:phase-wise} to \ref{sec:applying-mm-on-pi} describe the construction of transition matrices, distributional modelling of PI across phases of an innings, and contextual factors such as home/away conditions and tournament-specific behaviour. Building on these results, Section~\ref{sec:strategy} develops a prescriptive framework that provides situation-specific Pressure Index targets and strategic guidelines for teams aiming to control run chases in Twenty20 cricket.

\section{Review of literature}\label{sec:literature}
Though the paper is heavily dependent on the Pressure Index for formulating a strategic approach towards successful run chases in the Twenty20 format of cricket, the work of \cite{bhattacharjee2016quantifying} and their subsequent extended works, to name a few, \cite{talukdar2018identifying} and \cite{bhattacharjee2018measuring} are not the only contributions in this domain of tactic building for run chases.

Some earlier works that were concerned with the progress of scoring in the second innings of limited overs matches are \cite{bailey2006predicting} and \cite{de2001applications} While \cite{bailey2006predicting} attempted to predict the final score of the second innings of a One-day match while the game was in progress, \cite{de2001applications} attempted to estimate the magnitude of victory in One-day matches. In both cases the authors used the Duckworth--Lewis resource table at the end of each over of the second innings of the match.

\cite{choudhury2007use} used Multilayer Perceptron Neural Networks to predict the outcome of a cricket tournament involving more than two teams. \cite{sankaranarayanan2014towards} developed a prediction system that takes in historical ODI cricket match data as well as the instantaneous state of a match and predicts the score at key points in the future, culminating in a prediction of victory or loss. Using factors of the current situation of a cricket match, \cite{lokhande2018live} attempted to predict the outcome through a two-stage process. In the first stage, the match situation and other cricketing attributes are used to predict the first innings score and then in the second stage the possibility of the team batting second achieving the target is estimated. \cite{asif2016play} presented a model for forecasting the outcomes of One-Day International cricket matches when the game is in progress using dynamic logistic regression. \cite{o2006impress} compared four different score prediction tools by extending them to a ball-by-ball prediction model for predicting the final score, based on the present state of the first innings of a match. A series of papers are available in the literature that used different machine learning tools for predicting either the target of the match or the winner of the match when the match is in progress, viz. \cite{pathak2016applications}, \cite{passi2018increased}, \cite{vistro2019cricket}, \cite{modekurti2020setting}. Another set of researchers worked on the development of the Win and Score Predictor (WASP). The purpose of WASP is to predict the score at the end of the first innings of a limited overs match given the current situation and to predict the probability of a successful run chase of the team batting second given the target. Some literature concerning this includes \cite{hogan2017devising}, \cite{singh2015score}, and \cite{nimmagadda2018cricket}.

Most of the available literature regarding cricket match situation analysis are directed towards predicting final scores, match outcomes, or win probabilities using traditional or recent statistical or data mining tools. The models applied and comprehensive methodologies developed by previous researchers are highly valuable for forecasting. But such models are either descriptive or predictive in nature. However, in this current study, the researchers tried to develop a prescriptive model. By combining a measure of quantifying match pressure with a complex third-order Markov model, the current research attempts to frame strategies for successful run chases in Twenty20 cricket. The approach provides an exclusive, real-time tool for coaches to manage the pressure during run-chase on an over-by-over basis, which differentiates the contribution of the paper from the existing body of literature. 

\subsection{Objectives}\label{sec:objectives}
The paper achieves the following objectives:
\begin{enumerate}
    \item To develop a Markov model to quantify whether teams have exhibited any set pattern in successful run chases over a prolonged period of time.
    \item To study the patterns of run chases and accordingly prepare strategies in order to keep teams batting second in control throughout run chases for future Twenty20 matches.
\end{enumerate}

\section{Concepts of Markov Chain}\label{sec:markovchain}
\subsection{Markov chain}\label{sec:markovchain1}

\begin{definition}
A random process $\{X_n : n \ge 0\}$ is said to be a Markov chain with state space $S$ if it satisfies the Markov condition, i.e., for any states $x_0$, $x_1$, $x_2$, \dots, $x_{n-1}$, $i$, $j \in S$,
\begin{equation}
\label{eq:mcdef}
\mathbb{P}(X_{n+1} = j \mid X_n = i, X_{n-1} = x_{n-1}, \dots, X_1 = x_1, X_0 = x_0) = \mathbb{P}(X_{n+1} = j \mid X_n = i)
\end{equation}
\end{definition}

It can be easily verified that the Markov condition is equivalent to the condition: for any states $x_1$, $x_2$, \dots, $x_{k-1}$, $i$, $j \in S$,
\begin{equation}
\label{eq:mcdef2}
\mathbb{P}(X_{n+1} = j \mid X_{n_k} = i, X_{n_{k-1}} = x_{k-1}, \dots, X_{n_1} = x_1) = \mathbb{P}(X_{n+1} = j \mid X_{n_k} = i)
\end{equation}
for any $n_1 < n_2 < \ldots < n_k \leq n$.

The evolution of the chain clearly depends on the probabilities $\mathbb{P}(X_{n+1} = j \mid X_{n_k} = i)$. In general, this depends on $i$, $j$ and $n$. We restrict our attention to the case where this probability is independent of $n$.

\begin{definition}
A transition matrix $\mathbf{P} = ((p_{ij}))$ is the matrix whose $(i,j)$th entry is given by
\begin{equation}
\label{eq:transdef1}
((p_{ij})) = \mathbb{P}(X_{1} = j \mid X_{0} = i).
\end{equation}
\end{definition}

It is clear that, being conditional probabilities, the $p_{ij}$ 's must satisfy
\begin{equation}
\label{eq:transdef2}
p_{ij} \geq 0 \quad \text{and} \quad \sum_{j \in S} p_{ij} = 1 \quad \text{for all} \quad i \in S.
\end{equation}

\subsection{Markov chains of higher order}\label{sec:markovchain2}
The setup mentioned in Section~\ref{sec:markovchain1}, where the probability of an event at any time point depends only on that of the immediate past, can be thought of as a first order Markov chain. We may encounter situations where the probability of an event at a certain time point (say, the current state) might depend not just on the immediate past but on several time points prior to the current state. These are considered to be higher order Markov chains.

\begin{definition}
A random process $\{X_n : n \ge 0\}$ is said to be a Markov chain of order $k$ if for any states $x_0$, $x_1$, $x_2$, \dots, $x_{n-1}$, $i$, $j \in S$,
\begin{equation}
\label{eq:highermcdef}
\mathbb{P}(X_{n+1} = j \mid X_n = i, \dots, X_0 = x_0) = \mathbb{P}(X_{n+1} = j \mid X_n = i, \dots, X_{n-k+1} = x_{n-k+1}).
\end{equation}
\end{definition}

For our purpose, consider $\{X_n : n \ge 1\}$ to be the sequence of Pressure Index values (see details in Section~\ref{sec:introtopi}) for a team batting second at the end of each over during a run chase in a Twenty20 match, i.e., $X_n$ is the Pressure Index value at the end of the $(n+1)$th over during a run chase in a Twenty20 match. We will see why such a sequence is a suitable choice for Markov modelling.

\section{Introduction to Pressure Index}\label{sec:introtopi}
\subsection{Definitions and Formulation}\label{sec:pi}
As mentioned earlier, the Pressure Index ($PI$) is a ball-by-ball measure which quantifies the pressure experienced by the team batting second. In Twenty20 cricket, the team batting second has to achieve a certain target before its resources are exhausted. Thus, the pressure of the team batting second reduces if runs are scored at a fast rate and if a considerable number of wickets are kept in hand until the target is reached. Based on this idea, the formula of the Pressure Index \cite{bhattacharjee2016quantifying} for teams batting second in Twenty20 matches is given by:
\begin{equation}
\label{eq:pi}
PI = \frac{CRRR}{IRRR} \times \frac{1}{2} \left[ e^{\frac{RU}{100}} + e^{\frac{\sum w_i}{11}} \right] 
\end{equation}

where $IRRR$ is the initial required run rate, i.e., if $T$ is the target to be scored by the team batting second in $B$ balls, then
\begin{equation}
\label{eq:irrr}
IRRR = \frac{T \times 6}{B}.
\end{equation}
$CRRR$ denotes the current required run rate at any point during the second innings. If $R'$ runs have been scored in $B'$ balls at some stage, then
\begin{equation}
\label{eq:crrr}
CRRR = \frac{(T - R') \times 6}{B - B'}.
\end{equation}

The ratio $\frac{CRRR}{IRRR}$ thus measures the progress of the team batting second during a run chase. A lower value of the ratio denotes that the team is supposedly ahead in the chase.

The term $\sum w_i$ denotes the sum of the weights of the wickets that have fallen until any stage of the innings. As wickets fall, the team's wicket strength deteriorates. For example, if a top order batter gets out, the wicket strength of the team drops more than when a lower order batter gets out. In order to account for the varying abilities of batters affecting their wicket strengths, wicket weights ($w_i$) as per \cite{lemmer2015method} are used instead of assigning equal weights to every wicket. The wicket weights for different batting positions, as assigned by \cite{lemmer2015method}, are given below.

\begin{table}[H]
    \centering
    \caption{Wicket weights assigned by Lemmer}
    \label{tab:lemmer}
\begin{tabular}{l c c c c c c c c c c c}
\toprule
Position ($i$) & 1 & 2 & 3 & 4 & 5 & 6 & 7 & 8 & 9 & 10 & 11 \\
\midrule
Wicket weight ($w_i$) & 1.30 & 1.35 & 1.40 & 1.45 & 1.38 & 1.18 & 0.98 & 0.79 & 0.59 & 0.39 & 0.19 \\
\bottomrule
\end{tabular}
\end{table}

Finally, $RU$ denotes the percentage of resources used by the team at any stage during the innings as per the Duckworth--Lewis table.

Several other formulae to calculate the Pressure Index are available but we will avoid using those in this study due to various limitations. One such formula, as suggested by \cite{shah2014pressure}, is as follows:
\begin{equation}
\label{eq:shahandshah}
PI = CI \times 100 + \left[\frac{wicket\, weight}{180} \times T \times \frac{Br}{B} \times \frac{Rr}{T}\right] 
\end{equation}
According to the above formula, the PI value will always be 100 at the start. After calculating PI values for a considerable number of matches using this formula, it was found that PI decreased when a wicket fell and when the number of runs scored in an over was less than the required run rate. This is completely unrealistic considering the very purpose of the formulation of the Pressure Index and hence we will not take into account this formula.

Two other formulae which were also suggested, but had significant limitations, are given below.
\begin{equation}
\label{eq:otherpi1}
PI_1 = \frac{CRRR}{IRRR} \times e^{\frac{\sum w_i}{11}} 
\end{equation}
\begin{equation}
\label{eq:otherpi2}
PI_2 = \frac{CRRR}{IRRR} \times e^{\frac{RU}{100}} 
\end{equation}

The factor $CI = \frac{CRRR}{IRRR}$ is expected to change after every ball. While $e^{\frac{\sum w_i}{11}}$ changes only if a wicket falls, $e^{\frac{RU}{100}}$ changes after every ball as $RU$ takes into consideration not only the number of wickets fallen but also the balls faced by a team. At the start of the second innings, the Pressure Index of a team is equal to 1 according to both $PI_1$ and $PI_2$ and it will fluctuate as the match progresses. If the team batting second is able to achieve the target, both $PI_1$ and $PI_2$ become 0. However, it seems that $PI_2$ is more sensitive compared to $PI_1$ because it depends on $RU$, which is expected to change after every ball throughout the innings.

Considering all these factors, the formula \eqref{eq:pi} seems to be the most commendable, and we will consider the same for further study in this paper.

\subsection{Operational Mechanism}\label{sec:workingofpi}
It is evident from \eqref{eq:pi} that the Pressure Index value of a team batting second remains at unity at the start of a run chase. With every delivery bowled, the value might change and it gradually reaches zero once the team has achieved its target. In case the team fails to do so, the PI value rises steeply at the later stages of the match. To better understand how the indices work, the example of two matches have been taken.

Consider the Twenty20 match played between South Africa and India at Newlands Cricket Ground, Cape Town on February 24, 2018. South Africa started slowly but cautiously, chasing 173, $45/1$ after 9 overs. Although the required run rate was creeping up, 9 wickets were still in hand and they were still in the game. But just when they started to accelerate, wickets began falling, which resulted in a halt in their scoring rate leading to a rapid increase in the Pressure Index. By the 18th over, SA completely lost their grip on the run chase and finally fell short by 7 runs even after scoring 45 runs off their last 3 overs. On the other hand, Australia successfully chased down a record 244 runs, just a week prior to that, at Auckland with the strategy of never letting the required run rate rise too high.

\begin{figure}[ht]  
    \centering
    \includegraphics[width=0.6\textwidth, height=9cm]{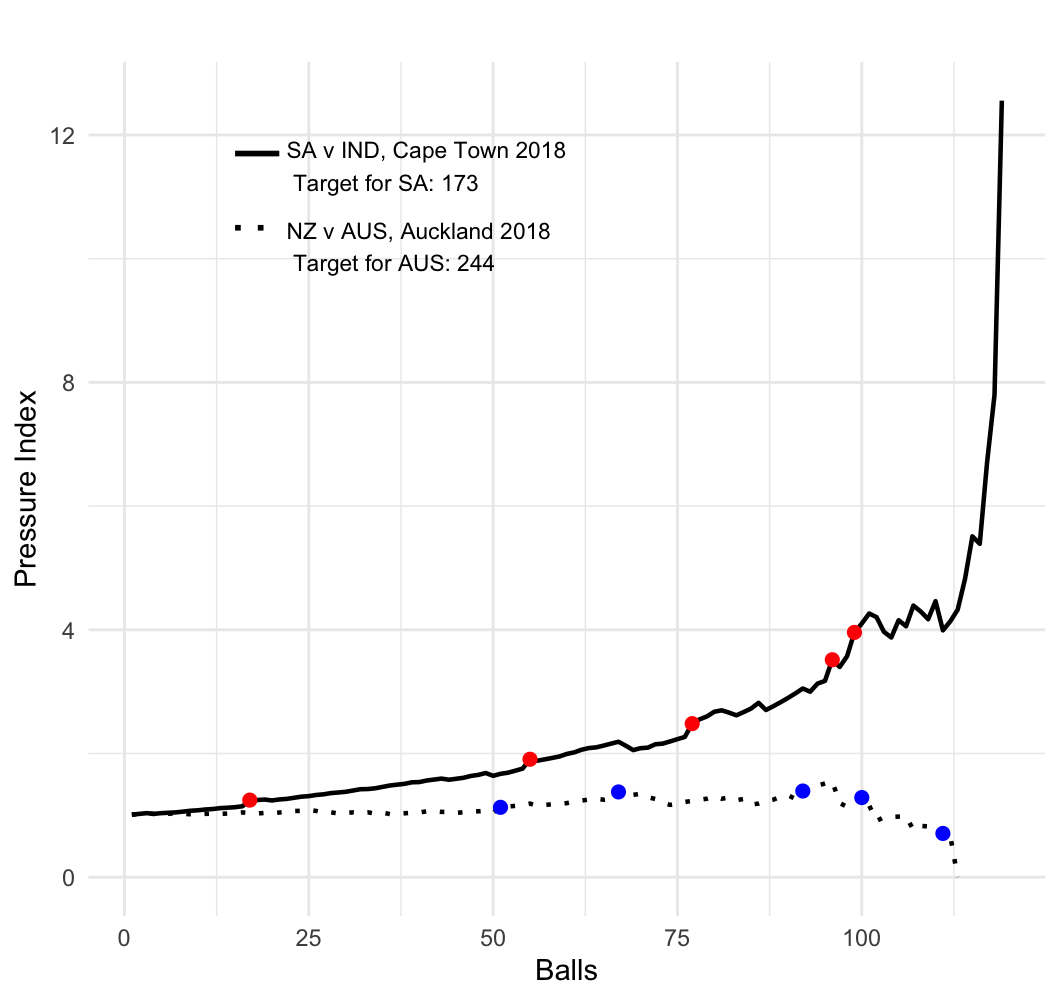}
    \caption{Pressure Curves for T20I 649 (NZ v AUS 2018) and T20I 655 (SA v IND 2018)}
    \label{fig:pressure-curves} 
    {\centering \small (Note: The dots in the pressure curves denote fall of wickets)\par}
\end{figure}

The urgency to score runs during a run chase increases as the game progresses, unless the target is too small to create any trouble for the team batting second. Also, scoring rates might halt when a wicket falls. The number of runs scored and wickets lost until any stage of the innings influence the scoring rate of the team in the next over. Furthermore, the target, the runs scored, and wickets lost until any stage of the innings determine the Pressure Index of the team at that stage. Thus, the Pressure Index at the end of an over during any stage of the innings depends on the Pressure Index of the team in the previous overs.

Thus, we can consider a sequence $\{X_n : n \ge 1\}$ where $X_n$ is the Pressure Index value at the end of the $(n+1)$th over during a run chase in a Twenty20 match. This leads to a setup where the concepts of Markov chains on the sequence $\{X_n : n \ge 1\}$ can be purposefully implemented.

\section{Methodology}\label{sec:methodology}

\subsection{Estimating the order of Markov chain}\label{sec:estimating-order}

We have defined our sequence to be the chronological $PI$ values at the end of each over of the second innings of a Twenty20 match but the number of preceding $PI$ values on which the $PI$ at a certain stage of the sequence depends is still unknown. In an ideal scenario, the runs scored and wickets lost in a certain over should technically depend on that in each of the previous overs in the innings. However, in order to avoid unnecessarily complex calculations and to achieve an optimum level of accuracy, one should try to find out the desired order of the Markov chain, i.e. the desired number of preceding $PI$ values on which the next $PI$ value in a sequence truly depends, that fits our data well and leads to a high enough precision for prediction purposes.

All men's Twenty20 matches in which the team batting second have either reached the target after batting for more than 18 overs or failed to reach the target and have lost the match by a margin not more than 10 runs have been considered for the study. Since the goal of our study is to prepare a strategy in support of teams batting second in Twenty20 matches, a certain section of the selected sample of matches will be helpful in replicating the patterns of successful run chases. The choice of such selection has also been made based on the idea of using a significantly large training sample in order to enhance the construction of the model.

Starting from the first ever officially recorded men's Twenty20 match on June 13, 2003 till October 15, 2025, a total of 8241 matches have been played where the outcomes have been either of the ones mentioned earlier. After removing matches in which complete ball-by-ball data are unavailable, we finally take into account a filtered set of 6537 matches\footnote{Data for these matches have been collected from \url{https://www.espncricinfo.com} and \url{https://cricsheet.org/matches/}.} to create a set of 6537 sequences of the form $\{s_n : n \ge 1\}$. The correct order of the Markov chain is then estimated using these sequences to proceed with further study.

For an order-$k$ Markov chain, the state space ($S$) has been defined to be the collection of all possible observed $PI$ values. Based on the probabilities of transitions,
$$
(s_{t-k}, s_{t-(k-1)}, \dots , s_{t-1}) \rightarrow s_t
$$
transition matrices have been constructed. For a match with $T$ overs, an order-$k$ Markov chain requires $(T-k)$ observations, and so each match contributes $(T-k)$ transitions. In our dataset, an average of $19.1$ overs have been played per match. For example, for an order-$3$ Markov chain, we have approximately $104592$ transitions\footnote{This count included repeated transitions as well. The number of unique transitions are listed in Table~\ref{tab:model-order-comparison}} in the pooled data. Note that each match is treated independently, so transitions do not cross match boundaries.

\begin{algorithm}[ht]
\caption{Construction of State Transitions from Pressure Index}
\label{alg:pi-transition}
\begin{algorithmic}[1] 

\For{each match $m$ in dataset}
    \State Extract PI values: $[PI_1, PI_2, \dots, PI_T]$
    
    \For{$t = k+1$ to $T$}
        \State previous state $\gets (PI_{t-k}, \dots, PI_{t-1})$
        \State next state $\gets PI_t$
        
        \State Record transition: previous state $\rightarrow$ next state
        \State Increment $count$[previous state, \ next state]
    \EndFor
\EndFor

\end{algorithmic}
\end{algorithm}

Negative $PI$ values are not of interest since they do not provide any inferential interpretation, so keeping them in analysis distorts distributional fits and transition matrices. Common statistical methods to handle them include censoring, truncation/flooring or redefining the state space etc. A simple alternative would be to set $PI=0$ immediately as the target in a run chase is achieved. This essentially results in censoring where we proceed our analysis with $PI^+ = max (PI, 0)$, instead of raw $PI$ which can take negative values. For brevity, we will denote $PI^+$ to be simply $PI$ throughout the study. It should be ensured that a previous state ending with $s_{t-1} =0$ cannot predict a non negative $PI$ value (see details in Section~\ref{sec:discretization}).

For a total of $M (=6537$) matches in our dataset,

\begin{equation}
\label{eq:N-transitions}
N_{transitions} = \sum_{m=1}^M (T_m - k)
\end{equation}

For an order-$k$ Markov chain, construct empirical probabilities
\begin{equation}
\label{eq:empirical-probabilities}
\mathbb{P}(s_t | s_{t-k}, s_{t-k+1}, \dots , s_{t-1}) = \frac{N(s_{t-k}, s_{t-k+1}, \dots , s_{t-1}, s_t)}{N(s_{t-k}, s_{t-(k-1)}, \dots , s_{t-1})},
\end{equation}
where  $N$ denotes the count of transitions in the training data. Log likelihood function of the Markov model is given by,

\begin{equation}
\label{eq:loglik}
L_k = \sum\limits_{i=1}^{N_{transitions}} \mathbb{P}(s_t ^{(i)} | \text{ previous state}^{(i)})
\end{equation}
Common model selection criteria include $AIC$ and $BIC$ which are defined as follows.

\begin{equation}
\label{eq:AIC}
AIC_k = 2m_k - 2L_k
\end{equation}
\begin{equation}
\label{eq:BIC}
BIC_k = m_k \operatorname{log} (N_{transitions}) - 2L_k
\end{equation}
where $m_k$ is the number of free parameters, i.e. the number of unique transitions in the transition matrix of the corresponding model.

\begin{table}[htbp]
    \centering
    \caption{Comaprison between models of different orders}
    \label{tab:model-order-comparison}
\begin{tabular}{c c c c c c c}
\toprule
Order & Observations & Parameters ($m_k$) & Log likelihood & AIC & BIC & P-to-O ratio \\
\midrule
1 & 116383 & 33034 & -439.8439 & 945755.9 & 1265017.7 & 0.2838 \\
2 & 115028 & 91573 & -372.0692 & 604431.9 & 1444466.3 & 0.7961 \\
3 & 110025 & 101829 & -2.7725 & 197590.4 & 950321.2 & 0.9255 \\
4 & 102127 & 103732 & -6.8451 & 190754.9 & 913820.6 & \textcolor{red}{1.0157} \\
5 & 90309 & 90321 & -5.2042 & 189364.6 & 870646.3 & \textcolor{red}{1.00001} \\
6 & 83826 & 83937 & -1.3862 & 187659.5 & 870115.9 & \textcolor{red}{1.001} \\
7 & 77348 & 77357 & -1.0332 & 164698.8 & 86747.6 & \textcolor{red}{1.00001} \\
\bottomrule
\end{tabular}
\end{table}

The entire data was treated as an $80\%-20\%$ split of training data and test data respectively and the results, from Table~\ref{tab:model-order-comparison}, show an order-$3$ model to have the highest log likelihood. Only models of order $3$ or lesser have a parameter-to-observations ratio lesser than 1, resulting in significant overfitting for higher order models. Although $AIC$ and $BIC$ keep on decreasing as order increases, they already approach a certain saturation point for an order-$3$ model. The extreme overfitting restricts the predictive accuracy for higher order models as well. As a result of a trade-off between the factors, we consider an order-$3$ model to proceed for further study. A pseudo-code for obtaining the optimal order of a Markov chain is given in Algorithm~\ref{alg:optimal-mc-order}.

An average batting pair (a partnership), in our dataset, lasts for $18.0126 \ (\approx18)$ balls during a Twenty20 innings. For every break in partnership, a wicket falls, so $RU$ and $\sum w_i$ change leading to a drastic change in $PI$. \cite{saikia2019cricket} explained how partnerships and $PI$ are related in Twenty20 matches. So heuristically, an order-$3$ Markov chain is expected to fit our dataset well. On the other hand, based on data from July 01, 2015\footnote{The choice of the date is based on the fact that the International Cricket Council (ICC) implemented major changes to rules in One Day Internationals, viz. elimination of batting powerplay, free hits for all kinds of no-balls etc, which brought a significant change in scoring rates in One Day Internationals post July 2015.} to October 15, 2025, an average batting pair lasts for $35.249 \ (\approx35)$ balls in a men's One Day International innings, in matches where a team batting second have either reached the target after batting for more than 45 overs or failed to reach the target and have lost the match by a margin not more than 15 runs. Moreover, a different Duckworth-Lewis table is used to extract $RU$ and different Wicket weights have been defined by \cite{article}, for One Day International (or in general $50$-over) matches. So, an order-$6$, instead of an order-$3$ Markov model, is expected to yield meaningful results for $50$-over formats.

\subsection{Discretization of Pressure Index}\label{sec:discretization}

After the transition matrix for an order-$3$ model has been obtained, any new sequence consisting of $k$ entries ($3 \leq k \leq T_j$), for some $j$-th match, can be inserted to predict the outcome of the unknown $(k+1)$th entry of the sequence. While a tuple of previous states is inserted, the model searches for occurrences of the exact set of previous states in the transition matrix. If multiple current states are observed corresponding to a tuple of previous states with non zero probabilities in the transition matrix, then the expected value of those observations is considered as the predicted outcome. However, if no occurrences of current states are found with non zero probability for a tuple of previous states, then a robust method to estimate the next $PI$, as described later in Section~\ref{sec:phase-wise} and Section~\ref{sec:applying-mm-on-pi}, would be to consider confidence intervals of $PI$ values and look to estimate a Pressure Index within a certain range with high confidence, using a distributional (here, gamma) fallback.

Equation~\ref{eq:pi} treats $PI$ values as continuous entities, hence taking values in an uncountable set. However in practice, an implicit discretization of $PI$ being rounded to two decimal places, i.e. a $0.01$ precision (or grid) has been performed.

The following example shows the outcome when $(1.49, 1.63, 1.48)$ is entered as the new sequence. In cricketing terms, this translates to a situation where the Pressure Index of a team batting second have been recorded as $1.49$, $1.63$ and $1.48$ at the end of three consecutive overs during the second innings of a Twenty20 match. Given the target and the number of overs the innings is into, we are interested in predicting the probable Pressure Index of the team at the end of the next over (see pseudo code in Algorithm~\ref{alg:predict-pi}). Table~\ref{tab:example1} provides the list of all possible $PI$ values a team has exhibited in the training sample after having a previous state $PI$ tuple ($1.49$, $1.63$, $1.48$) along with the corresponding probabilities and the resulting expected $PI$.

\begin{table}[H]
    \centering
    \caption{Expected outcomes of $PI$ values based on previous states}
    \label{tab:example1}
    \begin{adjustbox}{max width=\textwidth}
\begin{tabular}{c c | c c | c c}
\toprule
Pressure Index & \shortstack{Probability of\\Occurance} &
Pressure Index & \shortstack{Probability of\\Occurance} &
Pressure Index & \shortstack{Probability of\\Occurance} \\
\midrule
0.00 & 0.010310341 & 0.01 & 0.020620683 & 0.02 & 0.005155171 \\
0.31 & 0.010310341 & 0.66 & 0.010310341 & 0.82 & 0.010310341 \\
0.88 & 0.010310341 & 1.21 & 0.010310341 & 1.25 & 0.010310341 \\
1.36 & 0.010310341 & 1.39 & 0.010310341 & 1.40 & 0.010310341 \\
1.42 & 0.005155171 & 1.43 & 0.010310341 & 1.46 & 0.010310341 \\
1.47 & 0.010310341 & 1.48 & 0.010310341 & 1.51 & 0.015465512 \\
1.52 & 0.015465512 & 1.54 & 0.015465512 & 1.56 & 0.005155171 \\
1.57 & 0.010310341 & 1.58 & 0.015465512 & 1.59 & 0.036086194 \\
1.60 & 0.018867925 & 1.61 & 0.020620683 & 1.62 & 0.020620683 \\
1.64 & 0.034333436 & 1.66 & 0.041241365 & 1.67 & 0.030931024 \\
1.68 & 0.041241365 & 1.69 & 0.020620683 & 1.70 & 0.015465512 \\
1.71 & 0.030931024 & 1.72 & 0.046396536 & 1.73 & 0.046396536 \\
1.74 & 0.025775853 & 1.75 & 0.030931024 & 1.76 & 0.056706877 \\
1.79 & 0.005155171 & 1.80 & 0.010310341 & 1.81 & 0.030931024 \\
1.82 & 0.020620683 & 1.84 & 0.030931024 & 1.86 & 0.041241365 \\
1.87 & 0.010310341 & 1.90 & 0.010310341 & 1.91 & 0.010310341 \\
1.94 & 0.003402413 & 1.95 & 0.005155171 & 2.00 & 0.010310341 \\
2.02 & 0.020620683 & 2.23 & 0.010310341 & 2.39 & 0.010310341 \\
\bottomrule
\end{tabular}
\end{adjustbox}
\end{table}

\noindent\textbf{Expected Pressure Index outcome: 1.594674} $\approx$ \textbf{1.59}

\medskip

This means that if a team faces a situation where they have recorded Pressure Index values of $1.49$, $1.63$, $1.48$ in that order at the end of three consecutive overs during the second innings of a Twenty20 match, then the expected Pressure Index value at the end of the next over would be approximately $1.59$.

Inspite of a large number of transitions, finding exact replication of a $3$-tuple of $PI$ values from the transition matrix might be difficult and in some cases, impossible. Table~\ref{tab:common-transitions} shows that no set of $4$ consecutive $PI$ values have occured more than $4$ times in the matrix. Even after using a $0.01$ grid, $98.86 \%$ of the unique transitions appear only once in the matrix, indicating severe sparsity. To address this, Laplace smoothing was implemented which did not yield satisfactory results due to high proportion of singleton occurances, extremely low proportion ($0.11 \%$) of reliable states, i.e. states with 10 or more observations, and only a $30.85 \%$ coverage on the test data.

\begin{table}[H]
    \centering
    \caption{Few common transitions in order-$3$ model with $0.01$ precision}
    \label{tab:common-transitions}
\begin{tabular}{c c c | c | c}
\toprule
$PI_{t-3}$ & $PI_{t-2}$ & $PI_{t-1}$ & $PI_t$ & No. of occurances \\
\midrule
1.02 & 1.04 & 1.10 & 1.15 & 4 \\
1.03 & 1.05 & 1.07 & 1.11 & 4 \\
0.97 & 0.98 & 0.95 & 1.01 & 3 \\
0.97 & 1.03 & 1.01 & 1.01 & 3 \\
0.98 & 0.99 & 0.96 & 0.97 & 3 \\
0.98 & 1.05 & 1.12 & 1.15 & 3 \\
0.99 & 0.94 & 0.98 & 1.07 & 3 \\
0.99 & 1.01 & 1.02 & 1.09 & 3 \\
0.99 & 1.03 & 1.10 & 1.16 & 3 \\
\bottomrule
\end{tabular}
\end{table}

As a robust alternative, several precisions ranging from $0.01$ to $0.5$ were tested (see Algorithm~\ref{alg:discretization}) and the results are shown in Table~\ref{tab:precision-comparisons}. Based on the least MAE and RMSE values and a trade-off between proportions of singleton occurances and coverage probabilities, which signifies coarseness of the data, a $0.1$ precision has been chosen for further study.

\begin{table}[H]
    \centering
    \caption{Comparison between models with various discretization levels}
    \label{tab:precision-comparisons}
\begin{tabular}{c c c c c c}
\toprule
Precision & $N_{transitions}$ & Singleton $\%$ & MAE & RMSE & Coverage $\%$ \\
\midrule
0.01 & 81519 & 98.86 & 1.155 & 4.344 & 30.854 \\
0.025 & 68576 & 86.02 & 0.904 & 4.308 & 78.679 \\
0.05 & 42445 & 79.31 & 0.878 & 4.296 & 84.404 \\
0.075 & 31198 & 58.64 & 0.758 & 4.263 & 93.244 \\
0.1 & 22884 & 37.44 & 0.671 & 4.096 & 97.248 \\
0.15 & 18465 & 31.20 & 0.672 & 4.122 & 97.692 \\
0.2 & 11181 & 29.88 & 0.672 & 4.131 & 97.931 \\
0.25 & 10336 & 28.44 & 0.674 & 4.131 & 98.048 \\
0.5 & 3458 & 16.36 & 0.673 & 4.135 & 99.414 \\
\bottomrule
\end{tabular}
\end{table}

\subsection{Phase-wise Markov models}\label{sec:phase-wise}

Although Section~\ref{sec:estimating-order} and Section~\ref{sec:discretization} provide insights into how Pressure Index values can be monitored throughout an innings, it emerges as quite a general approach irrespective of a team's specific patterns during run chases in certain situations. For a better understanding, one might be interested in studying the patterns of where a team's Pressure Index lie at different stages during a successful run chase in Twenty20 cricket.

\begin{table}[H]
    \centering
    \caption{Phase-wise Marginal Distributional Analysis of Pressure Index}
    \label{tab:phase-pi-marginal-distr}
\begin{tabular}{l | c c c c c c}
\toprule
Phase & Mean $PI$ & Median $PI$ & $\text{SD}_{PI}$ & $25^{th}$ percentile & $75^{th}$ percentile & IQR \\
\midrule
Powerplay & 1.323 & 1.301 & 0.218 & 1.17 & 1.46 & 0.29 \\
Middle Overs & 1.74 & 1.692 & 0.412 & 1.45 & 1.99 & 0.54 \\
Death Overs & 2.859 & 2.641 & 1.488 & 1.77 & 3.68 & 1.91 \\
\bottomrule
\end{tabular}
\end{table}

Table~\ref{tab:phase-pi-marginal-distr} shows significant behavioral differences in $PI$ values during Powerplay (Overs 1 to 6), Middle Overs (Overs 7 to 16) and Death Overs (Overs 17 to 20) of innings during several successful run chases (comprising 4083 out of the 6537 matches from our initial training set). The choice of these specific phases are quite common in practice and are often found in literature for various phase-wise analysis of a Twenty20 innings. Variance of the marginal distribution of Pressure Index being the largest at Death Overs is in terms with the fact that uncertainity in a run chase generally increases as the innings progresses. Kolmogorov-Smirnov tests have been then performed over a few known distributions, and AIC and BIC have been observed (see Algorithm~\ref{alg:phase-wise}) to determine which known distribution fits the data corresponding to $PI$ values the best\footnote{The values corresponding to AIC, BIC and Kolmogorv-Smirnov tests for each of the three distributions in the tables have been calculated using \texttt{fitdistrplus} and \texttt{MASS} packages in R.}.

\begin{table}[H]
    \centering
    \caption{Results for Pressure Index in Powerplay}
    \label{tab:fit-powerplay}
\begin{tabular}{l c c c}
\toprule
Criteria & Gamma & Exponential & Weibull \\
\midrule
$AIC$ & $-1267.731$ & $6438.255$ & $-718.434$ \\
$BIC$ & $-1264.717$ & $7119.283$ & $-724.320$ \\
K-S test statistic $(D)$ & $0.026477$ & $0.48476$ & $0.055296$ \\
\bottomrule
\end{tabular}
\end{table}

\begin{table}[H]
    \centering
    \caption{Results for Pressure Index in Middle Overs}
    \label{tab:fit-middle}
\begin{tabular}{l c c c}
\toprule
Criteria & Gamma & Exponential & Weibull \\
\midrule
$AIC$ & $632.815$ & $6192.033$ & $965.436$ \\
$BIC$ & $653.670$ & $6186.024$ & $971.667$ \\
K-S test statistic $(D)$ & $0.013766$ & $0.400127$ & $0.047228$ \\
\bottomrule
\end{tabular}
\end{table}

\begin{table}[H]
    \centering
    \caption{Results for Pressure Index in Death Overs}
    \label{tab:fit-death}
\begin{tabular}{l c c c}
\toprule
Criteria & Gamma & Exponential & Weibull \\
\midrule
$AIC$ & $1977.545$ & $7019.481$ & $3118.065$ \\
$BIC$ & $1992.033$ & $7439.283$ & $3126.974$ \\
K-S test statistic $(D)$ & $0.017849$ & $0.32761$ & $0.06927$ \\
\bottomrule
\end{tabular}
\end{table}

Lower AIC, BIC and Kolmogorov-Smirnov test statistic values denote that the gamma distribution fits well for the $PI$ values corresponding to each section of the data. This gives us an idea of how the $PI$ values at three different stages of an innings are distributed\footnote{The density plots corresponding to the three phases of a Twenty20 innings have been constructed using \texttt{fitdistrplus} and \texttt{ggplot2} packages in R.}.

\begin{table}[htbp]
    \centering
    \caption{Estimated parameters of Gamma distribution}
    \label{tab:gamma-parameters}
\begin{tabular}{l c c}
\toprule
Phases & Estimated shape parameter & Estimated rate parameter \\
\midrule
Powerplay & $38.276$ & $28.931$ \\
Middle Overs & $18.447$ & $10.62$ \\
Death Overs & $3.667$ & $1.286$ \\
\bottomrule
\end{tabular}
\end{table}

Using the estimated gamma parameters in Table~\ref{tab:gamma-parameters}, the phase means are obtained and the variance increases markedly towards the death overs. Practically, this means $PI$ is relatively concentrated near its phase mean during the Powerplay and becomes more dispersed as innings progresses. We exploit this in two ways within our strategy: (a) when empirical Markov transitions are sparse for a given tuple, the fitted gamma provides a phase-specific fallback predictive distribution for $PI$ which is then used to compute expected PI and confidence intervals, and (b) the gamma quantiles are used to map $PI$ ranges into the suitable zones (see details in Section~\ref{sec:strategy}). Thus, besides descriptive overlays, the gamma fits provide phase-specific probabilistic approximations that increase coverage when transition data are missing and frame a actionable strategy by linking percentiles to tactical thresholds.

\begin{figure}[htbp]  
    \centering
    \includegraphics[width=0.8\textwidth, height=7cm]{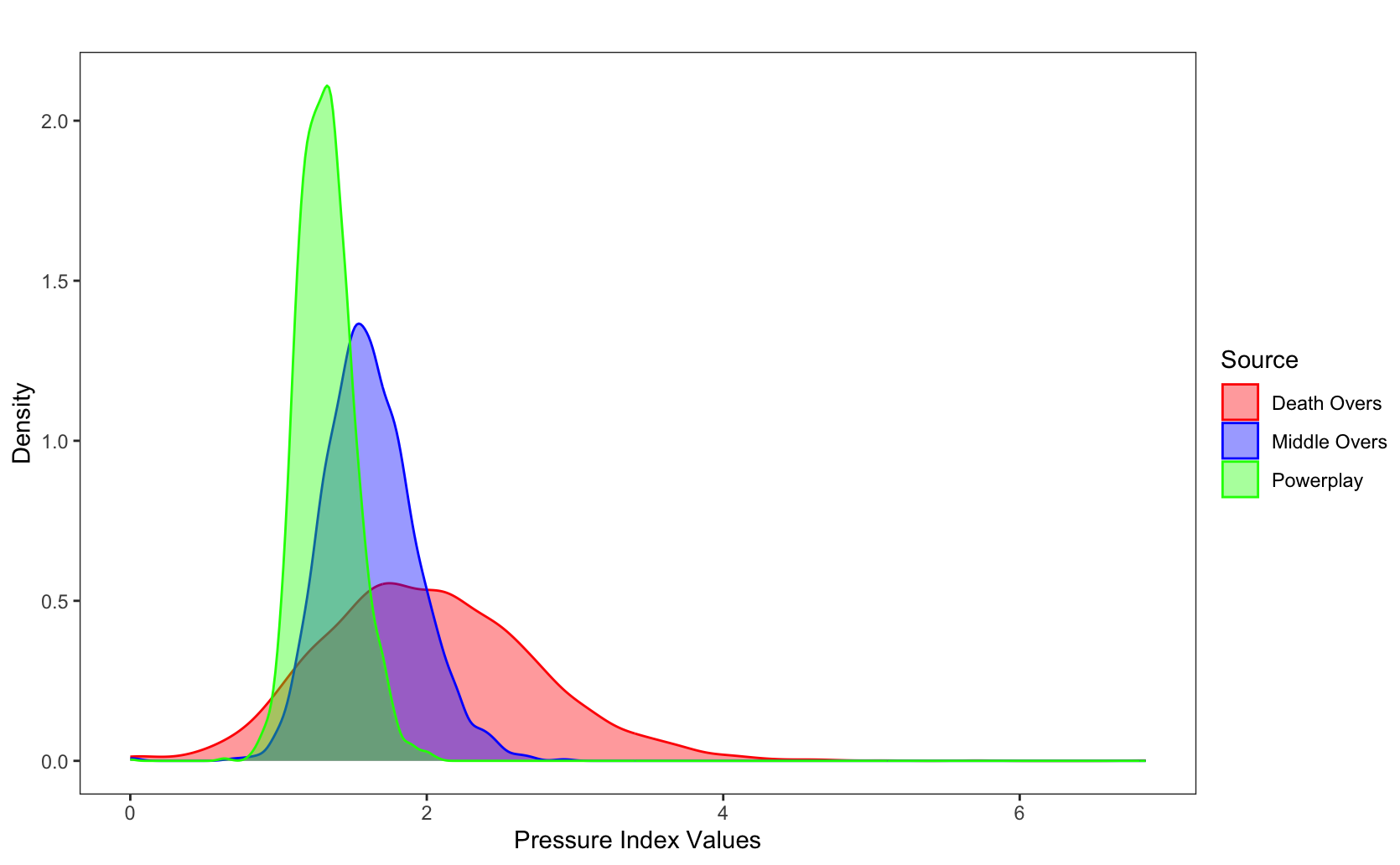}
    \caption{Density plots of phase-wise Pressure Index}
    \label{fig2} 
\end{figure}

The Kolmogorov-Smirnov statistic $D$ is reported only for descriptive comparison. Since parameters are estimated from the same data, classical K-S p-values are invalid. A valid goodness-of-fit assessment is therefore performed using a parametric bootstrap (see Algorithm~\ref{alg:bootstrap-ks}), which refits the model to bootstrap samples to obtain 
the empirical null distribution of $D$.

Separate transition matrices corresponding to the three different phases of an innings can be constructed based on the process described in earlier sections and outcomes resulting from a tuple consisting of three $PI$ values will then be obtained based on the transition matrix corresponding to the specific section of the innings only. Results in Table~\ref{tab:global-vs-phase} clearly imply that phase-wise models exhibit better performance than a single generalised global model in terms of lesser MAE and RMSE.

\begin{table}[htbp]
    \centering
    \caption{Comparison of Global vs Phase-wise Markov models}
    \label{tab:global-vs-phase}
\begin{tabular}{l | c c c c}
\toprule
& \multicolumn{2}{c}{MAE} & \multicolumn{2}{c}{RMSE} \\
\cmidrule(lr){2-3} \cmidrule(lr){4-5}
Phase & Global & Phase-wise & Global & Phase-wise \\
\midrule
Powerplay & 0.944 & 0.912 & 6.23 & 5.74 \\
Middle Overs & 0.877 & 0.209 & 3.03 & 0.292 \\
Death Overs & 0.173 & 0.151 & 0.316 & 0.197 \\
\bottomrule
\end{tabular}
\end{table}

Models based on the three-phase partition of the innings is expected to avoid the overgeneralization of teams' strategies in run chases caused by the global model (see details in Section~\ref{sec:applying-mm-on-pi}). 

\subsection{Contextual Factors in Pressure Index Dynamics}\label{sec:other-factors}

\subsubsection{Discrepancies in Pressure Index for home and away teams}\label{sec:home-vs-away}

Besides the three-phase partitioning of a Twenty20 innings, several external factors like home ground advantage, winning the toss ahead of a match, team strengthening due to player-player coalition or player-coach coalition etc. might play crucial roles in determining teams' strategy for a successful run chase in Twenty20 matches. In our dataset, around $12.52 \%$ of the matches have been played at neutral venues (e.g. South Africa vs India at Perth Stadium on October 30, 2022) or at venues which are home grounds for both the participating teams (e.g. Bangladesh vs Bangladesh A at Sher-E Bangla National Stadium, Mirpur on December 11, 2013). Excluding this section of matches, home teams have won $53.93 \%$ of the games, which rises up to $57.42 \%$ if we consider Twenty20 International matches only.

\begin{table}[ht!]
    \centering
    \caption{Win rates induced by phase-wise $PI$ thresholds}
    \label{tab:phase-wise-hva}
\begin{tabular}{l c c c c}
\toprule
Phases & $PI$ thresholds & \shortstack{Home team win $\%$\\($H \%$)} & \shortstack{Away team win $\%$\\($A \%$)} & $\Delta = H-A$ \\
\midrule
Powerplay & $[0, 0.5)$ & $100$ & $100$ & $0 \%$ \\
          & $[0.5, 1)$ & $73.7$ & $62.2$ & $+ 11.6 \%$ \\
          & $[1, 1.5)$ & $70.6$ & $61.8$ & $+ 8.8 \%$ \\
          & $[1.5, 2)$ & $68.2$ & $58.5$ & $+ 9.7 \%$ \\
          & $[2, 2.5)$ & $42.7$ & $36.6$ & $+ 6.1 \%$ \\
          & $[2.5, 3)$ & $14.6$ & $11.1$ & $+ 3.5 \%$ \\
\midrule
Middle Overs & $[0, 0.5)$ & $100$ & $100$ & $0 \%$ \\
             & $[0.5, 1)$ & $100$ & $98.4$ & $+ 1.6 \%$ \\
             & $[1, 1.5)$ & $75.7$ & $70.7$ & $+ 5.0 \%$ \\
             & $[1.5, 2)$ & $51.3$ & $43.8$ & $+ 7.6 \%$ \\
             & $[2, 2.5)$ & $28.8$ & $22.1$ & $+ 6.6 \%$ \\
             & $[2.5, 3)$ & $12.4$ & $12.1$ & $+ 0.3 \%$ \\
             & $[3, 3.5)$ & $5.9$ & $2.1$ & $+ 3.9 \%$ \\
             & $[3.5, 4)$ & $1.2$ & $1.5$ & $- 0.3 \%$ \\
             & $[4, 4.5)$ & $1.2$ & $0$ & $+ 1.2 \%$ \\
             & $[4.5, 5)$ & $0$ & $0$ & $0 \%$ \\
\midrule
Death Overs  & $[0, 0.5)$ & $100$ & $100$ & $0 \%$ \\
             & $[0.5, 1)$ & $100$ & $98.8$ & $+ 1.2 \%$ \\
             & $[1, 1.5)$ & $94.4$ & $94.3$ & $+ 0.1 \%$ \\
             & $[1.5, 2)$ & $85.7$ & $81.6$ & $+ 4.1 \%$ \\
             & $[2, 2.5)$ & $67.5$ & $60.2$ & $+ 7.3 \%$ \\
             & $[2.5, 3)$ & $55.7$ & $47.5$ & $+ 8.2 \%$ \\
             & $[3, 3.5)$ & $36.4$ & $14.7$ & $+ 21.7 \%$ \\
             & $[3.5, 4)$ & $25.9$ & $11.1$ & $+ 14.8 \%$ \\
             & $[4, 4.5)$ & $8.8$ & $0.6$ & $+ 8.2 \%$ \\
             & $[4.5, 5)$ & $3.5$ & $0.9$ & $+ 2.6 \%$ \\
\bottomrule
\end{tabular}
\end{table}

$\Delta$ takes positive values throughout Table~\ref{tab:phase-wise-hva} for most choices of $PI$ thresholds across all three phases of an innings. Moreover teams having $100 \%$ win rates when $PI \in [0, 0.5)$, irrespective of the phase of an innings, is in terms with the fact that if the Pressure Curve during a run chase reaches values less than 0.5, then the chasing team is highly likely to win, according to \cite{saikia2019cricket} (details in Section~\ref{sec:strategy}). A negative $\Delta$ for $PI \in [3.5, 4)$ during Middle Overs is due to rare occurances of outliers (e.g. Royal Challengers Bangalore vs Chennai Super Kings at M Chinnaswamy Stadium on April 25, 2018), supported on lower sample sizes for matches with extremely high $PI$ values. Another notable discrepancy from Table~\ref{tab:phase-wise-hva} is the rise of $A$ from $0$ to $0.6$ and $0.9$ for $PI \in [4, 4.5)$ and $[4.5, 5)$ respectively, which has been also caused due to a rare outlier: Royal Challengers Bangalore vs Kolkata Knight Riders at M Chinnaswamy Stadium on April 05, 2019.

Besides the abundance of positive $\Delta$ values throughout Table~\ref{tab:phase-wise-hva}, the differences ($\Delta$) are significantly large for higher $PI$ values as the innings progresses, i.e. towards the Death Overs. A deeper partition of the innings, based on Overs $16$ to $19$ (see Table~\ref{tab:death-overs-hva}) clearly shows teams playing in home conditions have historically shown an edge in handling pressure, towards the final stages of run chases, more than those playing in away conditions. Note that the choice of Overs $16$ to $19$ instead of taking the Pressure Index at the end of Over $20$ into account, is based on the fact that $PI$ values at the end of Over $20$ is largely concentrated at $0$ for successfully completed run chases and around too high values for unsuccessful run chases, leading to an extremely heavy-tailed distribution which has no useful interpretation for predictive purposes.

\begin{table}[H]
    \centering
    \caption{Win rates induced by Over-wise $PI$ thresholds}
    \label{tab:death-overs-hva}
\begin{tabular}{l c c c c}
\toprule
Over & $PI$ thresholds & \shortstack{Home team win $\%$\\($H \%$)} & \shortstack{Away team win $\%$\\($A \%$)} & $\Delta = H-A$ \\
\midrule
End of $16^{th}$ over & $[0, 0.5)$ & $100$ & $100$ & $0 \%$ \\
                      & $[0.5, 1)$ & $100$ & $100$ & $+ 0 \%$ \\
                      & $[1, 1.5)$ & $94.2$ & $94.1$ & $+ 0.1 \%$ \\
                      & $[1.5, 2)$ & $86.6$ & $85.6$ & $+ 1 \%$ \\
                      & $[2, 2.5)$ & $70.2$ & $56.9$ & $+ 13.2 \%$ \\
                      & $[2.5, 3)$ & $38.8$ & $29.4$ & $+ 9.4 \%$ \\
                      & $[3, 3.5)$ & $18.8$ & $7.4$ & $+ 11.4 \%$ \\
                      & $[3.5, 4)$ & $4.3$ & $3.5$ & $+ 0.8 \%$ \\
                      & $[4, 4.5)$ & $3$ & $0$ & $+ 3 \%$ \\
                      & $[4.5, 5)$ & $0$ & $0$ & $0 \%$ \\
\midrule
End of $17^{th}$ over & $[0, 0.5)$ & $100$ & $100$ & $0 \%$ \\
                      & $[0.5, 1)$ & $100$ & $100$ & $0 \%$ \\
                      & $[1, 1.5)$ & $94.7$ & $91.3$ & $+ 3.4 \%$ \\
                      & $[1.5, 2)$ & $88.6$ & $87.2$ & $+ 1.4 \%$ \\
                      & $[2, 2.5)$ & $70.9$ & $62.7$ & $+ 8.2 \%$ \\
                      & $[2.5, 3)$ & $56.9$ & $45.7$ & $+ 11.2 \%$ \\
                      & $[3, 3.5)$ & $29.3$ & $6.9$ & $+ 22.4 \%$ \\
                      & $[3.5, 4)$ & $11.4$ & $4.1$ & $+ 7.3 \%$ \\
                      & $[4, 4.5)$ & $5.6$ & $0$ & $+ 5.6 \%$ \\
                      & $[4.5, 5)$ & $2.5$ & $0$ & $+ 2.5 \%$ \\
\midrule
End of $18^{th}$ over  & $[0, 0.5)$ & $100$ & $100$ & $0 \%$ \\
                       & $[0.5, 1)$ & $98.8$ & $93.4$ & $+ 5.4 \%$ \\
                       & $[1, 1.5)$ & $90.5$ & $89.9$ & $+ 0.6 \%$ \\
                       & $[1.5, 2)$ & $82.9$ & $78.1$ & $+ 4.8 \%$ \\
                       & $[2, 2.5)$ & $76.6$ & $66.7$ & $+ 9.9 \%$ \\
                       & $[2.5, 3)$ & $60.7$ & $51.6$ & $+ 9.1 \%$ \\
                       & $[3, 3.5)$ & $34.8$ & $20$ & $+ 14.8 \%$ \\
                       & $[3.5, 4)$ & $27.3$ & $14.2$ & $+ 13.1 \%$ \\
                       & $[4, 4.5)$ & $10.5$ & $0.6$ & $+ 9.9 \%$ \\
                       & $[4.5, 5)$ & $3.6$ & $0.9$ & $+ 2.7 \%$ \\
\midrule
End of $19^{th}$ over  & $[0, 0.5)$ & $100$ & $100$ & $0 \%$ \\
                       & $[0.5, 1)$ & $100$ & $91.8$ & $+ 8.2 \%$ \\
                       & $[1, 1.5)$ & $89.4$ & $87.3$ & $+ 2.1 \%$ \\
                       & $[1.5, 2)$ & $73.3$ & $66.8$ & $+ 6.5 \%$ \\
                       & $[2, 2.5)$ & $60.1$ & $53.2$ & $+ 6.9 \%$ \\
                       & $[2.5, 3)$ & $61.5$ & $47.5$ & $+ 14 \%$ \\
                       & $[3, 3.5)$ & $41.2$ & $28.6$ & $+ 12.6 \%$ \\
                       & $[3.5, 4)$ & $25$ & $22.2$ & $+ 2.8 \%$ \\
                       & $[4, 4.5)$ & $8.8$ & $0.6$ & $+ 8.2 \%$ \\
                       & $[4.5, 5)$ & $3.7$ & $0.9$ & $+ 2.8 \%$ \\
\bottomrule
\end{tabular}
\end{table}

\subsubsection{Tournament-specific Model performance}\label{sec:tournament-specific}

Owing to the variability of playing conditions, competitive intensity and other deciding factors, it could be of interest to check how the model behaves in terms of predictive performance for different tournaments.

Denote \textit{Coverage percentage} to be the percentage of times the actual Pressure Index lies within the $95 \%$ confidence interval of the predicted Pressure Index and \textit{Markov Usage percentage} to be the percentage of predicted $PI$ values that use the Markov transition matrix (actual observed data) over the Gamma distribution fallback (statistical approximation). Table~\ref{tab:tournament-results} shows the predictive performance of a few popular tournaments based on the Markov model (see details in Appendix~\ref{sec:tables} for the entire list).

\begin{table}[H]
\centering
\caption{Tournament-wise performance of the Markov model}
\label{tab:tournament-results}
\begin{tabular}{lcccc}
\toprule
Competition & MAE & RMSE & Coverage $\%$ & Markov Usage $\%$ \\
\midrule
IPL & 0.8885 & 7.8697 & 84.16 & 95.86 \\
T20 Blast & 6.4472 & 14.1590 & 83.37 & 95.90 \\
Bangladesh Premier League & 1.0662 & 4.5498 & 80.58 & 96.12 \\
CPL & 5.5460 & 8.9030 & 84.31 & 95.10 \\
PSL & 0.6081 & 3.6784 & 84.53 & 95.93 \\
Syed Mushtaq Ali Trophy & 1.6174 & 3.3080 & 84.97 & 95.66 \\
SA20 League & 6.3580 & 13.7701 & 84.16 & 93.70 \\
\bottomrule
\end{tabular}
\end{table}

From Table~\ref{tab:tournament-results}, one can note a significant difference in MAE and RMSE between leagues played in Asia (IPL, PSL, Syed Mushtaq Ali Trophy etc. has lower MAE and RMSE) and those played outside Asia (T20 Blast, CPL, SA20 League etc. has higher MAE and RMSE). This discrepancy might occur due to various external factors like kind of pitches, dimensions of grounds etc. leading to differences in scoring rates between tournaments and so, it may be of interest to construct Markov models based on datasets of a specific tournament (equivalently, league or competition) and check how they perform in terms of predictive responses for an independent dataset, i.e. a different tournament.

\begin{table}[H]
    \centering
    \caption{Predictions on IPL, BBL and PSL based on the Markov model for IPL data}
    \label{tab:ipl-to-others}
    \setlength{\tabcolsep}{3pt}
    \scriptsize
    \begin{adjustbox}{max width=\textwidth}
\begin{tabular}{l | c c c c | c c c c | c c c c}
\toprule
& \multicolumn{4}{c}{IPL} & \multicolumn{4}{c}{BBL} & \multicolumn{4}{c}{PSL} \\
\cmidrule(lr){2-5} \cmidrule(lr){6-9} \cmidrule(lr){10-13}
Phase & MAE & RMSE & Coverage & \shortstack{Markov\\Usage} & MAE & RMSE & Coverage & \shortstack{Markov\\Usage} & MAE & RMSE & Coverage & \shortstack{Markov\\Usage} \\
\midrule
Powerplay & 0.102 & 0.129 & 86.5 & 100 & 2.209 & 8.236 & 85.4 & 100 & 2.112 & 7.144 & 83.5 & 98.8 \\
Middle Overs & 0.165 & 0.219 & 93.7 & 100 & 1.178 & 0.293 & 89.4 & 99.8 & 2.18 & 0.363 & 92.4 & 99.7 \\
Death Overs & 0.296 & 0.856 & 83.2 & 98.6 & 0.313 & 1.913 & 78.5 & 97.5 & 0.342 & 1.78 & 77.5 & 94.8 \\
\bottomrule
\end{tabular}
\end{adjustbox}
\end{table}

\begin{table}[hbtp]
    \centering
    \caption{Predictions on IPL, T20Is and CPL based on the Markov model for BBL data}
    \label{tab:bbl-to-others}
    \setlength{\tabcolsep}{3pt}
    \scriptsize
    \begin{adjustbox}{max width=\textwidth}
\begin{tabular}{l | c c c c | c c c c | c c c c}
\toprule
& \multicolumn{4}{c}{IPL} & \multicolumn{4}{c}{T20Is} & \multicolumn{4}{c}{CPL} \\
\cmidrule(lr){2-5} \cmidrule(lr){6-9} \cmidrule(lr){10-13}
Phase & MAE & RMSE & Coverage & \shortstack{Markov\\Usage} & MAE & RMSE & Coverage & \shortstack{Markov\\Usage} & MAE & RMSE & Coverage & \shortstack{Markov\\Usage} \\
\midrule
Powerplay & 1.097 & 6.124 & 82.7 & 100 & 2.115 & 8.151 & 78.4 & 99.6 & 1.104 & 7.132 & 81.6 & 100 \\
Middle Overs & 1.183 & 1.307 & 92.5 & 99.5 & 0.247 & 0.396 & 89.5 & 98.3 & 1.205 & 0.355 & 90.6 & 99.3 \\
Death Overs & 0.895 & 0.856 & 70.1 & 92.1 & 1.355 & 1.996 & 63.1 & 91.2 & 1.914 & 0.699 & 79.9 & 82.7 \\
\bottomrule
\end{tabular}
\end{adjustbox}
\end{table}

Table~\ref{tab:ipl-to-others} and Table~\ref{tab:bbl-to-others} clearly show several examples of how predictive performances can vary depending on contextual factors involved in a tournament.

\subsection{Applying the Markov model on Pressure Index}\label{sec:applying-mm-on-pi}

Based on our findings so far, we can predict the Pressure Index values with $0.1$ precision, along with suitable confidence intervals supported by phase-wise Gamma distribution fits, at the end of every over during the second innings of a Twenty20 match. Consider the Twenty20 match between Pakistan and West Indies at National Stadium, Karachi on April 03, 2018. Pakistan won the match by 8 wickets while chasing a target of 154 runs and our model captured a fair amount of the actual $PI$ values concentrated around the corresponding predicted $PI$ values.

\begin{table}[H]
    \centering
    \caption{Pressure Index of Pakistan: vs West Indies, April 03, 2018}
    \label{tab:pak-vs-wi}
    \begin{adjustbox}{max width=\textwidth}
\begin{tabular}{c | c c | c c c | c c c}
\toprule
&  &  & \multicolumn{3}{c}{$0.01$ precision} & \multicolumn{3}{c}{$0.1$ precision} \\
\cmidrule(lr){4-6} \cmidrule(lr){7-9}
Overs & \shortstack{Cumulative\\Runs} & \shortstack{Cumulative\\Wickets} & Actual PI & Predicted PI & $95 \%$ C.I & Actual PI & Predicted PI & $95 \%$ C.I \\
\midrule
1  & 5  & 0 & 1.04 & - & - & 1.0 & - & - \\
2  & 15  & 0 & 1.04 & - & - & 1.0 & - & -\\
3  & 32  & 0 & 0.99 & - & - & 1.0 & - & - \\
4  & 51  & 0 & 0.91 & 1.09 & [0.94, 1.24] & 0.9 & 1.0 & [0.9, 1.2] \\
5  & 57  & 0 & 0.93 & 0.82 & [0.67, 0.97] & 0.9 & 0.8 & [0.7, 1.0] \\
6  & 62  & 1 & 1.04 & 0.82 & [0.67, 0.97] & 1.0 & 0.8 & [0.7, 1.0] \\
7  & 71  & 1 & 1.03 & 0.92 & [0.78, 1.06] & 1.0 & 0.9 & [0.7, 1.1] \\
8  & 79  & 1 & 1.03 & 0.94 & [0.80, 1.08] & 1.0 & 0.9 & [0.7, 1.1] \\
9  & 86  & 1 & 1.05 & 1.05 & [0.86, 1.24] & 1.1 & 1.0 & [0.9, 1.2] \\
10 & 89  & 1 & 1.13 & 1.13 & [0.97, 1.29] & 1.1 & 1.1 & [1.0, 1.3] \\
11 & 98  & 1 & 1.11 & 1.08 & [0.94, 1.22] & 1.1 & 1.1 & [1.0, 1.3] \\
12 & 107  & 1 & 1.08 & 1.09 & [0.95, 1.23] & 1.1 & 1.1 & [1.0, 1.3] \\
13 & 114 & 2 & 1.15 & 1.14 & [0.97, 1.31] & 1.2 & 1.1 & [1.0, 1.3] \\
14 & 120 & 2 & 1.17 & 1.18 & [1.03, 1.32] & 1.2 & 1.2 & [1.1, 1.4] \\
15 & 132 & 2 & 0.83 & 0.83 & [0.67, 0.99] & 1.0 & 0.8 & [0.7, 1.0] \\
16 & 139 & 2 & 0.62 & 0.63 & [0.47, 0.79] & 1.0 & 0.6 & [0.5, 0.8] \\
17 & 154 & 2 & 0 & 0.06 & [0.00, 0.27] & 0 & 0 & [0.0, 0.2] \\
\bottomrule
\end{tabular}
\end{adjustbox}
\end{table}

From Table~\ref{tab:pak-vs-wi}, it can be observed that 12 out of the 14 predictions have been successful, in the sense that the actual $PI$ lie within the $95 \%$ confidence interval of the corresponding predicted $PI$, for the model with $0.01$ precision, which improves to 13 out of 14 successful predictions for the model with $0.1$ precision. For some more assessment, we look at a rather closely contested Twenty20 match played between Chennai Super Kings and Delhi Capitals at Dubai International Cricket Stadium on October 10, 2021. Inspite of the Pressure Index never reducing below $1$ throughout the second innings, Chennai Super Kings chased down a target of 173 runs in the 20th over and our model performed fairly well throughout.

\begin{table}[H]
    \centering
    \caption{Pressure Index of Chennai Super Kings: vs Delhi Capitals, October 10, 2021}
    \label{tab:csk-vs-dc}
    \begin{adjustbox}{max width=\textwidth}
\begin{tabular}{c | c c | c c c | c c c}
\toprule
&  &  & \multicolumn{3}{c}{$0.01$ precision} & \multicolumn{3}{c}{$0.1$ precision} \\
\cmidrule(lr){4-6} \cmidrule(lr){7-9}
Overs & \shortstack{Cumulative\\Runs} & \shortstack{Cumulative\\Wickets} & Actual PI & Predicted PI & $95 \%$ C.I & Actual PI & Predicted PI & $95 \%$ C.I \\
\midrule
1  & 8  & 1 & 1.08 & - & - & 1.0 & - & - \\
2  & 16  & 1 & 1.11 & - & - & 1.1 & - & -\\
3  & 20  & 1 & 1.24 & - & - & 1.2 & - & - \\
4  & 34  & 1 & 1.13 & 1.15 & [1.01, 1.29] & 1.1 & 1.2 & [1.1, 1.4] \\
5  & 39  & 1 & 1.19 & 1.18 & [1.04, 1.33] & 1.2 & 1.2 & [1.1, 1.4] \\
6  & 59  & 1 & 1.14 & 1.21 & [0.99, 1.28] & 1.1 & 1.2 & [1.1, 1.4] \\
7  & 64  & 1 & 1.21 & 1.23 & [1.00, 1.31] & 1.2 & 1.2 & [1.1, 1.4] \\
8  & 68  & 1 & 1.30 & 1.21 & [0.98, 1.29] & 1.3 & 1.2 & [1.1, 1.4] \\
9  & 75  & 1 & 1.36 & 1.35 & [1.19, 1.51] & 1.4 & 1.4 & [1.3, 1.6] \\
10 & 81  & 1 & 1.45 & 1.46 & [1.31, 1.61] & 1.5 & 1.5 & [1.4, 1.7] \\
11 & 94  & 1 & 1.43 & 1.49 & [1.35, 1.63] & 1.4 & 1.5 & [1.4, 1.7] \\
12 & 99  & 1 & 1.56 & 1.46 & [1.31, 1.61] & 1.6 & 1.5 & [1.4, 1.7] \\
13 & 111 & 1 & 1.51 & 1.67 & [1.52, 1.82] & 1.5 & 1.7 & [1.6, 1.9] \\
14 & 117 & 3 & 1.69 & 1.56 & [1.40, 1.72] & 1.7 & 1.6 & [1.5, 1.8] \\
15 & 121 & 4 & 2.29 & 1.97 & [0.83, 1.11] & 2.3 & 2.0 & [1.9, 2.2] \\
16 & 129 & 4 & 2.49 & 2.46 & [2.31, 2.61] & 2.5 & 2.5 & [2.4, 2.7] \\
17 & 138 & 4 & 2.68 & 2.52 & [2.34, 2.70] & 2.7 & 2.5 & [2.4, 2.7] \\
18 & 149 & 4 & 2.85 & 2.79 & [2.62, 2.97] & 2.9 & 2.8 & [2.7, 3.0] \\
19 & 160 & 5 & 3.39 & 3.61 & [3.42, 3.80] & 3.4 & 3.6 & [3.5, 3.8] \\
20 & 173 & 6 & 0 & 4.40 & [4.21, 4.59] & 0 & 4.4 & [4.3, 4.6] \\
\bottomrule
\end{tabular}
\end{adjustbox}
\end{table}

Teams win only $19.1 \%$ of matches while entering the 20th over of a run chase with a Pressure Index higher than $3.5$, which is even lower ($13.6 \%$) for matches played until 2021. Based on this fact, the run chase in Table~\ref{tab:csk-vs-dc} is closer to an outlier than to a general representative of the data, inspite of which the model exhibits 13 out of 16 successful predictions on both a $0.01$ scale and $0.1$ scale of precision, until the start of the last over.

The above examples show how the indices can be monitored throughout the innings and how continuously revising the probabilities of keeping the Pressure Index within a suitable range (see details in Section~\ref{sec:strategy}) helps a team remain in control during a run chase in Twenty20 cricket. A pseudo-code to calculate $PI$ during run chases, from ball by ball data, is given in Algorithm~\ref{alg:pi-calculate-bbb}.

\section{Strategy for controlled run chases using the Markov model}\label{sec:strategy}

It has been observed from extensive studies that if the Pressure Curve during a run chase reaches values less than $0.5$, then the chasing team is highly likely to win. Tables~\ref{tab:phase-wise-hva} and \ref{tab:death-overs-hva} have also supported this fact in our dataset. From \cite{saikia2019cricket}, it is understood that
\begin{equation}
\label{eq:saikia-prob0to5}
\mathbb{P}(\text{match won by team batting second} \mid 0 < PI < 0.5) = 1,
\end{equation} 
and for the different subintervals of $PI$ values between $[0.5, 3.5]$, viz. $[0.5, \lambda_{i_1}]$, $[\lambda_{i_1}, \lambda_{i_2}]$, $\dots$, $[\lambda_{i_m}, 3.5]$, the values of 
\[
\mathbb{P}(\text{match won by team batting second} \mid \lambda_i < PI < \lambda_j)
\]
generally start with $1$ but keep on decreasing until they eventually reach $0$. An immediate strategy for teams batting second in Twenty20 matches would be to try to bring down the Pressure Index as low as possible, and effectively below $0.5$, at some stage during the innings in order to guarantee a win. However, based on our findings in Section~\ref{sec:methodology}, it might be of interest to obtain situation-specific thresholds for Pressure Index which enhances the probability of successfully completing a run chase.

\begin{table}[H]
    \centering
    \caption{Situation-specific Pressure Index target recommendations}
    \label{tab:pi-thresholds-recommendations}
    \begin{adjustbox}{max width=\textwidth}
\begin{tabular}{c | c | c c | c c}
\toprule
&  & \multicolumn{2}{c}{Home teams} & \multicolumn{2}{c}{Away teams} \\
\cmidrule(lr){3-4} \cmidrule(lr){5-6}
Phases & $PI$ Thresholds & Win Rate & Recommendation & Win Rate & Recommendation \\
\midrule
Powerplay    & $[0, 0.5)$ & $100 \%$ & Target Zone & $100 \%$ & Target Zone \\
             & $[0.5, 1)$ & $73.7 \%$ & Acceptable & $62.2 \%$ & Acceptable \\
             & $[1, 1.5)$ & $42.7 \%$ & Risky & $36.6 \%$ & Avoid \\
             & $[1.5, 2.5)$ & $13.9 \%$ & Avoid & $10.5 \%$ & Avoid \\
             & $[2.5, \infty)$ & $0 \%$ & Avoid & $0 \%$ & Avoid \\
\midrule
Middle Overs & $[0, 0.5)$ & $100 \%$ & Target Zone & $100 \%$ & Target Zone \\
             & $[0.5, 1)$ & $100 \%$ & Target Zone & $98.4 \%$ & Target Zone \\
             & $[1, 1.5)$ & $75.7 \%$ & Acceptable & $70.7 \%$ & Acceptable \\
             & $[1.5, 2.5)$ & $43.6 \%$ & Risky & $35.9 \%$ & Avoid \\
             & $[2.5, \infty)$ & $6.9 \%$ & Avoid & $3.7 \%$ & Avoid \\
\midrule
Death Overs  & $[0, 0.5)$ & $100 \%$ & Target Zone & $100 \%$ & Target Zone \\
             & $[0.5, 1)$ & $100 \%$ & Target Zone & $96.8 \%$ & Target Zone \\
             & $[1, 1.5)$ & $87.3 \%$ & Acceptable & $70.2 \%$ & Acceptable \\
             & $[1.5, 2.5)$ & $75.9 \%$ & Acceptable & $68.2 \%$ & Acceptable \\
             & $[2.5, \infty)$ & $11.2 \%$ & Avoid & $8.1 \%$ & Avoid \\
\bottomrule
\end{tabular}
\end{adjustbox}
\end{table}

The recommendations for a team on which Pressure Index thresholds one should try to approach during a run chase, have been categorized as four zones, viz. Target Zone, Acceptable Zone, Risky Zone and Avoid Zone, in order of diminishing favourability based on historical win rates and predictive behaviour of the phase-wise Markov model. The terminology 'Zone' has been used to keep parity with 'Pressure Zone' as defined by \cite{saikia2019cricket}.
 
Note that the 'Target zone' and 'Acceptable zone' of $PI$ thresholds tend to enlarge as the innings progresses. The unbounded intervals $[2.5, \infty)$ in Table~\ref{tab:pi-thresholds-recommendations} are practically not unbounded (see \cite{saikia2019cricket} for methods to control abruptly rising Pressure Index towards the end of an innings during unsuccessful run chases) and should not be a matter of interest for our purpose.

\begin{table}[H]
\centering
\caption{Over-wise predictive performance of the Markov model}
\label{tab:mae-rmse-overwise}
\begin{tabular}{c | c c c c c}
\toprule
Overs & Mean Actual $PI$ & Mean Predicted $PI$ & Coverage & MAE & RMSE \\
\midrule
 4  & 1.13 & 1.17 & 99.7 $\%$ & 1.162 & 6.189 \\
 5  & 1.21 & 1.24 & 99.2 $\%$ & 1.031 & 5.220 \\
 6  & 1.32 & 1.32 & 98.4 $\%$ & 0.470 & 5.244 \\
 \cdashline{1-6}
 7  & 1.41 & 1.45 & 98.1 $\%$ & 0.184 & 0.294 \\
 8  & 1.52 & 1.54 & 94.4 $\%$ & 0.190 & 0.349 \\
 9  & 1.63 & 1.68 & 92.6 $\%$ & 0.207 & 0.137\\
10  & 1.75 & 1.84 & 88.1 $\%$ & 0.233 & 0.186\\
11  & 1.88 & 1.94 & 82.0 $\%$ & 0.298 & 0.548 \\
12  & 2.03 & 2.12 & 89.2 $\%$ & 0.502 & 0.690 \\
13  & 2.19 & 2.32 & 88.5 $\%$ & 0.713 & 1.900 \\
14  & 2.38 & 2.39 & 84.3 $\%$ & 1.270 & 1.460 \\
15  & 2.59 & 3.17 & 81.3 $\%$ & 1.960 & 1.880 \\
16  & 2.89 & 3.25 & 84.8 $\%$ & 1.217 & 6.920 \\
\cdashline{1-6}
17  & 3.33 & 3.41 & 80.6 $\%$ & 1.170 & 8.377 \\
18  & 3.62 & 3.71 & 79.8 $\%$ & 1.700 & 5.300\\
19  & 3.90 & 3.74 & 76.5 $\%$ & 2.683 & 8.582 \\
\bottomrule
\end{tabular}
\end{table}

Table~\ref{tab:mae-rmse-overwise} shows the mean of actual and predicted Pressure Index values fairly match, along with significantly high coverage percentages for every over of a Twenty20 innings for our test dataset. The over-wise MAE and RMSE are observed to be in parity with phase-wise MAE and RMSE shown in Table~\ref{tab:global-vs-phase}. On the other hand, relatively higher deviations and smaller coverage probabilities towards the end of the innings are expected to be caused due to the concentration around $0$ and too high $PI$ values, denoting higher uncertainity of run chases during Death Overs.

\begin{definition}
Brier score is a strictly proper scoring rule\footnote{A scoring rule $S$ is said to be proper, relative to a convex class $\mathcal{F}$ of probability measures, if its expected score is minimized when the forecasted distribution matches the distribution of the observation, i.e. $\mathbb{E}_{Y \sim Q} [S(Q,Y)] \leq \mathbb{E}_{Y \sim Q} [S(F,Y)]$ for all $F,Q \in \mathcal{F}$. It is strictly proper if this equation holds with equality if and only if $F=Q$.} that measures the accuracy of probabilistic predictions. For unidimensional predictions, it is equivalent to the mean squared error as applied to predicted probabilities.
\begin{equation}
\label{eq:brier-score}
Brier\ Score = \frac{\sum(\text{Predicted probability} - \text{Actual outcome})^2}{\text{Total number of predictions}}
\end{equation}
\end{definition}

\begin{definition}
Expected Calibration Error (ECE) is a measure of quantification of the discrepancy between predicted probabilities and empirical outcome frequencies by aggregating calibration error over a partition of the probability space.
\begin{equation}
\label{eq:ece}
ECE = \sum_{k=1}^{\text{Number of bins}} \left( \frac{\text{Number of predictions in } k^{\text{th}} \text{ bin}}{\text{Total number of predictions}} \right) \ |p_k - o_k| \\
\end{equation}
where,\qquad
\begin{minipage}{0.1\linewidth}
\begin{align*}
p_k &= \text{Predicted probability in } k^{\text{th}} \text{ bin},\\
o_k &= \text{Observed frequency in } k^{\text{th}} \text{ bin}.
\end{align*}
\end{minipage}
\end{definition}

\begin{table}[htbp]
\centering
\caption{Calibration of predicted probabilities for $PI \in [0, 0.5)$}
\label{tab:calibration-0to0.5}
\begin{tabular}{c | c c c}
\toprule
Probability bins & Predicted probability & Observed frequency & Calibration error \\
\midrule
0.0-0.1 & 0.0438 & 0.0498 & 0.006 \\
0.1-0.2 & 0.159 & 0.145 & 0.014 \\
0.2-0.3 & 0.280 & 0.221 & 0.059 \\
0.3-0.4 & 0.359 & 0.374 & 0.015 \\
0.4-0.5 & 0.421 & 0.436 & 0.015 \\
0.5-0.6 & 0.588 & 0.537 & 0.051 \\
0.6-0.7 & 0.609 & 0.673 & 0.064 \\
0.7-0.8 & 0.747 & 0.766 & 0.019 \\
0.8-0.9 & 0.852 & 0.886 & 0.034 \\
0.9-1.0 & 0.915 & 0.982 & 0.067 \\
\bottomrule
\end{tabular}
\end{table}

\begin{table}[htbp]
\centering
\caption{Calibration of predicted probabilities for $PI \in [0.5, 3.5)$}
\label{tab:calibration-0.5to3.5}
\begin{tabular}{c | c c c}
\toprule
Probability bins & Predicted probability & Observed frequency & Calibration error \\
\midrule
0.0-0.1 & 0.0451 & 0.0448 & 0.0003 \\
0.1-0.2 & 0.167 & 0.152 & 0.015 \\
0.2-0.3 & 0.272 & 0.254 & 0.018 \\
0.3-0.4 & 0.354 & 0.340 & 0.014 \\
0.4-0.5 & 0.487 & 0.448 & 0.039 \\
0.5-0.6 & 0.575 & 0.604 & 0.029 \\
0.6-0.7 & 0.622 & 0.661 & 0.039 \\
0.7-0.8 & 0.749 & 0.783 & 0.034 \\
0.8-0.9 & 0.855 & 0.871 & 0.015 \\
0.9-1.0 & 0.914 & 0.972 & 0.058 \\
\bottomrule
\end{tabular}
\end{table}

\begin{table}[H]
\centering
\caption{Brier Score and ECE for calibration of probabilities in Table~\ref{tab:calibration-0to0.5} and Table~\ref{tab:calibration-0.5to3.5}}
\label{tab:brier-ece}
\begin{tabular}{c | c c}
\toprule
Pressure Index & Brier Score & Expected Calibration Error \\
\midrule
$PI \in [0, 0.5)$ & 0.0184 & 0.00072 \\
$PI \in [0.5, 3.5)$ & 0.0245 & 0.00031 \\
\bottomrule
\end{tabular}
\end{table}

Tables~\ref{tab:calibration-0to0.5} and \ref{tab:calibration-0.5to3.5} show fairly good predictive performance for probabilities of Pressure Index lying in $[0, 0.5)$ and $[0.5, 3.5)$, along with very low Brier Score and ECE (see Table~\ref{tab:brier-ece}), which strengthens the motivation to frame a desired strategy from a coach or team's perspective during run chases in Twenty20 cricket.

\section{Conclusion}\label{conclusion}

The work focuses on the use of higher-order Markov chains to analyse the extent to which the current value of pressure, quantified by the Pressure Index ($PI$), influences future situations of run chases in limited overs cricket, primarily Twenty20 cricket. Previously, no study has addressed the extent to which the current match situation influences future situations. This approach demonstrates that a third order Markov chain is capable of capturing the complex, time-dependent patterns of $PI$ values, where the current pressure state depends on the immediate previous three values of the Pressure Index.

The current methodology utilises a vast dataset of $6537$ Twenty20 matches spanning over a period of $22$ years to build robust transition matrices, developing an immediate and real-time actionable tool. Crucially, this model is the first to use Pressure Index transitions not only for prediction of outcomes (viz. WASP by \cite{hogan2017devising} or Duckworth-Lewis methods), but also paves the way for understanding the strategy of successful run chases by analysing historical patterns for a large dataset.

The concept of computing the expected value of the Pressure Index for the next over, given the current match situation, allows the coaching staff and players to set immediate mini-targets so that the PI values remain between 0.5 to 1, (below 0.5 may not always be practically achievable), or at times within $1.5$ (see Table~\ref{tab:pi-thresholds-recommendations}) as early as possible \textemdash{} which the analysis depicts as a reasonable condition for a successful run-chase. This dynamic and real-time outcome of the study, with over-by-over tactical guidelines, is an innovative approach to the study.

\section{Direction of future research}\label{sec:future-research}

Tracing Pressure curves (\cite{bhattacharjee2016quantifying}) and revising Pressure Index values with the help of the proposed Markov model helps to build a strategy to keep teams batting second in Twenty20 matches in control throughout the chase. However, several potential developments to the model can be considered as part of future research:

\begin{enumerate}
        \renewcommand{\labelenumi}{\roman{enumi}.}
    \item Different teams have different patterns of a run chase. For example, teams with greater batting depths might score at a faster rate even after the fall of a wicket and still succeed but the same strategy might not work for teams with lesser batting depths. Therefore, errors are expected to occur when using the transition probabilities corresponding to one team to predict Pressure Index values for a team with a different composition.
    
    \item Besides the effects of contextual factors like home ground advantage and tournament-specific models discussed in Section~\ref{sec:other-factors}, external factors like venues, psychological factors like player-player coalition or player-coach coalition etc. might also play key roles in run chases in Twenty20 matches. Modification of the model to incorporate such external factors, using small sample methodologies, might increase the accuracy while predicting Pressure Index values.
    
    \item The model could be improved by considering the differences in the qualities and strengths of the teams involved in a match. Apart from the external factors, the ability of teams winning close contests, strengths of oppositions etc. might affect a team's performance. For example, a Pressure Index of say, $1.42$, at some stage of the innings against a stronger opposition may have a different impact compared to that against a weaker opposition. In matches between a top-ranked team and a lower-ranked team, the former is expected to capitalize towards the end of the match even from less favorable situations, primarily due to the significant difference in experience possessed by the two teams.
    
    \item The formula in \ref{eq:pi} restricts the usage of Pressure Index to second innings only. Certain variance stabilizing transformations (suggested by \cite{saikia2019cricket}), where $\frac{CRRR}{IRRR}$ is raised to a power $(\frac{CRRR}{IRRR})^\alpha$ for some suitable choice of $\alpha$, help to restrict the abrupt rise in $PI$ values at later stages of the innings during unsuccessful run chases. Such transformations can be applied in constructing a similar Markov model if one intends to use it for prediction purposes for teams batting first as well.
    
    \item The methodology used in this study could be extended to other formats of limited overs cricket, for example One-Day Internationals, with suitable modifications.
\end{enumerate}

\bibliographystyle{apalike} 
\bibliography{references}

\appendix

\section{Appendix}\label{sec:appendix}

\subsection{Necessary tables}\label{sec:tables}

\begin{table}[H]
\centering
\caption{Predictive performance of the Markov model for some tournaments/series}
\label{tab:tournament-appendix}
\begin{tabular}{p{0.7cm} l | c c c c}
\toprule
& \textbf{Competition} & \textbf{MAE} & \textbf{RMSE} & \textbf{Coverage} & \textbf{Markov Usage} \\
\midrule
\multirow{23}{*}{\rotatebox{90}{\textbf{Tournaments}}}
& T20 Internationals & 1.7749 & 4.0912 & 82.43\% & 93.64\% \\
& Abu Dhabi T20 League & 7.1149 & 13.9755 & 82.35\% & 94.12\% \\
& Africa Regional T20 Cup & 6.2063 & 16.5576 & 82.55\% & 95.22\% \\
& Bangabandhu T20 Cup & 3.1105 & 4.7792 & 87.96\% & 95.24\% \\
& Bangladesh Premier League & 1.0662 & 3.5498 & 80.58\% & 96.12\% \\
& BBL & 1.8664 & 3.1107 & 84.43\% & 95.92\% \\
& CPL & 5.5460 & 8.9030 & 84.31\% & 95.10\% \\
& CSA T20 Challenge & 3.3286 & 4.4902 & 84.16\% & 95.52\% \\
& Dhaka T20 League & 4.8326 & 6.5975 & 81.73\% & 94.31\% \\
& Emerging Teams Asia Cup & 1.9141 & 3.7764 & 84.05\% & 93.25\% \\
& ILT20 & 1.0259 & 2.7928 & 83.86\% & 95.48\% \\
& IPL & 0.8885 & 7.8697 & 84.16\% & 95.86\% \\
& Lanka Premier League & 3.9791 & 9.9558 & 84.62\% & 95.78\% \\
& Major League Cricket & 2.0930 & 5.1130 & 83.61\% & 95.48\% \\
& National T20 Cup & 5.5214 & 8.1287 & 84.90\% & 95.41\% \\
& PSL & 0.6081 & 3.6784 & 84.53\% & 95.93\% \\
& Ram Slam T20 Challenge & 2.1653 & 5.0489 & 81.12\% & 96.76\% \\
& SA20 League & 6.3580 & 13.7701 & 84.16\% & 93.70\% \\
& SLC Twenty-20 Tournament & 4.3099 & 9.4595 & 82.05\% & 96.46\% \\
& Super Smash & 1.5733 & 2.8123 & 84.49\% & 95.06\% \\
& Super T20 Provincial Tournament & 5.429 & 12.769 & 92.80\% & 93.07\% \\
& Syed Mushtaq Ali Trophy & 1.6174 & 3.3080 & 84.97\% & 95.66\% \\
& T20 Blast & 6.4472 & 14.1590 & 83.37\% & 95.90\% \\
\\[-0.5em]

\cdashline{2-6}
\\[-0.5em]

\multirow{14}{*}{\rotatebox{90}{\textbf{Few randomly picked series}}}
& Afghanistan A in Oman T20s & 7.8297 & 5.8597 & 78.79\% & 96.97\% \\
& Australia v Sri Lanka 2022 & 0.4093 & 1.1497 & 84.29\% & 92.86\% \\
& BBL 2023 & 0.7681 & 2.7041 & 87.59\% & 97.19\% \\
& CPL 2023 & 4.1487 & 3.0939 & 84.75\% & 94.92\% \\
& CPL 2024 & 3.9750 & 4.8647 & 84.25\% & 95.00\% \\
& ICC T20 WC Warm-up 2022 & 1.3045 & 3.0810 & 85.02\% & 87.50\% \\
& Ireland A in Bangladesh T20Is & 0.2406 & 0.4900 & 100.00\% & 100.00\% \\
& New Zealand v England 2019 & 1.0183 & 6.7465 & 88.24\% & 88.24\% \\
& SA20 League 2024 & 5.3580 & 7.7701 & 78.16\% & 93.70\% \\
& SL2020 & 4.0565 & 6.1665 & 79.49\% & 94.12\% \\
& South Africa A in Zimbabwe T20Is & 1.0273 & 4.6591 & 83.33\% & 89.58\% \\
& Syed Mushtaq Ali Trophy 2018–19 & 0.9208 & 1.9607 & 83.78\% & 99.10\% \\
& West Indies A in Nepal T20s & 1.1591 & 5.0552 & 84.67\% & 87.65\% \\
\bottomrule
\end{tabular}
\end{table}

\subsection{Pseudo codes for useful Algorithms}\label{sec:algorithms}

\begin{algorithm}[htbp]
\caption{Determine optimal Markov chain order}
\label{alg:optimal-mc-order}
\begin{algorithmic}[1]
\Require Dataset of $M$ matches with PI sequences
\Require Maximum order $k_{max}$
\Ensure Optimal order $k^*$

\State Split data into 80\% training and 20\% test sets
\For{$k = 1$ to $k_{max}$}
    \State Initialize transition count matrix $N^{(k)}$
    \State $N_{transitions} \leftarrow \sum_{m=1}^{M} (T_m - k)$ \Comment{Equation~\ref{eq:N-transitions}}
    \For{each match $m$ in training set}
        \For{$t = k+1$ to $T_m$}
            \State Extract previous state: $(s_{t-k}, \ldots, s_{t-1})$
            \State Extract current state: $s_t$
            \State Increment $N^{(k)}[(s_{t-k}, \ldots, s_{t-1}), s_t]$
        \EndFor
    \EndFor
    \State Calculate $m_k$ = number of unique transitions
    \State Calculate log-likelihood: $L_k \leftarrow \sum_{i=1}^{N_{transitions}} \log P(s_t^{(i)} | \text{previous state}^{(i)})$
    \State Calculate $AIC_k \leftarrow 2m_k - 2L_k$ \Comment{Equation~\ref{eq:AIC}}
    \State Calculate $BIC_k \leftarrow m_k \log(N_{transitions}) - 2L_k$ \Comment{Equation~\ref{eq:BIC}}
    \State Calculate parameter-to-observation ratio: $\rho_k \leftarrow \frac{m_k}{N_{transitions}}$
\EndFor
\State Select $k^* \leftarrow \arg\min_k \{AIC_k, BIC_k : \rho_k < 1\}$

\Return $k^*$
\end{algorithmic}
\end{algorithm}

\begin{algorithm}[htbp]
\caption{Discretize Pressure Index values}
\label{alg:discretization}
\begin{algorithmic}[1]
\Require Continuous PI values $\{PI_1, PI_2, \ldots, PI_n\}$
\Require Precision level $\delta$
\Ensure Discretized PI values $\{PI'_1, PI'_2, \ldots, PI'_n\}$

\For{each $PI_i$ in dataset}
    \If{$PI_i < 0$}
        \State $PI'_i \leftarrow 0$ \Comment{Censoring negative values}
    \Else
        \State $PI'_i \leftarrow \delta \times \lfloor \frac{PI_i}{\delta} + 0.5 \rfloor$ \Comment{Round to nearest $\delta$}
    \EndIf
\EndFor

\Return $\{PI'_1, PI'_2, \ldots, PI'_n\}$
\end{algorithmic}
\end{algorithm}

\begin{algorithm}[htbp]
\caption{Predict next Pressure Index value}
\label{alg:predict-pi}
\begin{algorithmic}[1]
\Require Transition matrix $\mathcal{T}$
\Require Previous states $(s_{t-k}, \ldots, s_{t-1})$
\Require Phase-specific Gamma parameters $(\alpha, \beta)$
\Require Confidence level $\gamma$
\Ensure Predicted PI value $\hat{s}_t$ with confidence interval $[\hat{s}_t^{lower}, \hat{s}_t^{upper}]$

\If{$(s_{t-k}, \ldots, s_{t-1})$ exists in $\mathcal{T}$}
    \State Extract all possible next states: $\{s_t^{(1)}, s_t^{(2)}, \ldots, s_t^{(n)}\}$
    \State Extract corresponding probabilities: $\{p_1, p_2, \ldots, p_n\}$
    \State Calculate expected value: $\hat{s}_t \leftarrow \sum_{i=1}^{n} p_i \cdot s_t^{(i)}$
\Else
    \State Calculate sum: $S \leftarrow \sum_{j=t-k}^{t-1} s_j$
    \State Search for states in $\mathcal{T}$ where $\sum_{j=t-k}^{t-1} s_j' = S$
    \If{matching states found}
        \State Calculate expected value from matching transitions
    \Else
        \State Use Gamma distribution fallback: $\hat{s}_t \leftarrow \frac{\alpha}{\beta}$ \Comment{Mean of Gamma}
    \EndIf
\EndIf
\State Calculate confidence interval using Gamma quantiles:
\State $\hat{s}_t^{lower} \leftarrow$ Gamma$^{-1}(\frac{1-\gamma}{2}; \alpha, \beta)$
\State $\hat{s}_t^{upper} \leftarrow$ Gamma$^{-1}(\frac{1+\gamma}{2}; \alpha, \beta)$

\Return $\hat{s}_t, [\hat{s}_t^{lower}, \hat{s}_t^{upper}]$
\end{algorithmic}
\end{algorithm}

\begin{algorithm}[htbp]
\caption{Fit phase-wise distributions to Pressure Index}
\label{alg:phase-wise}
\begin{algorithmic}[1]
\Require PI values for phase $\phi \in \{\text{Powerplay, Middle, Death}\}$
\Require Distribution families $\mathcal{D} = \{\text{Gamma, Exponential, Weibull}\}$
\Ensure Best-fit distribution parameters for phase $\phi$

\For{each distribution $d \in \mathcal{D}$}
    \State Estimate parameters $\theta_d$ using Maximum Likelihood Estimation
    \State Calculate log-likelihood: $L_d$
    \State Calculate $AIC_d \leftarrow 2|\theta_d| - 2L_d$
    \State Calculate $BIC_d \leftarrow |\theta_d| \log(n) - 2L_d$
    \State Perform Kolmogorov-Smirnov test: $(D_d, p_d) \leftarrow \text{KS-test}(PI, d, \theta_d)$
\EndFor
\State Select $d^* \leftarrow \arg\min_d \{AIC_d\}$ subject to $p_{d^*} > 0.05$

\Return Distribution $d^*$ with parameters $\theta_{d^*}$
\end{algorithmic}
\end{algorithm}

\begin{algorithm}[htbp]
\caption{Parametric Bootstrap for Kolmogorov-Smirnov Goodness-of-Fit}
\label{alg:bootstrap-ks}
\begin{algorithmic}[1]
\Require Observed data $x_1,\dots,x_n$; distribution family $\mathcal{D}$ (e.g., Gamma)
\Ensure K--S statistic $D_{\text{obs}}$ and bootstrap p-value $\hat{p}$

\State Fit $\mathcal{D}$ to observed data to obtain parameter estimate $\hat{\theta}$
\State Compute $D_{\text{obs}} = \sup_x | F_n(x) - F(x;\hat{\theta}) |$

\State Choose bootstrap size $B$

\For{$b = 1$ to $B$}
    \State Generate sample $x_1^{(b)},\dots,x_n^{(b)} \sim \mathcal{D}(\hat{\theta})$
    \State Fit $\mathcal{D}$ to bootstrap sample to obtain $\hat{\theta}^{(b)}$
    \State Compute $D^{(b)} = \sup_x |F_n^{(b)}(x) - F(x;\hat{\theta}^{(b)})|$
\EndFor

\State Compute bootstrap p-value:
$\hat{p} = \frac{1}{B} \sum_{b=1}^B \mathbf{1}\{D^{(b)} \ge D_{\text{obs}}\}$

\Return $D_{\text{obs}}, \hat{p}$
\end{algorithmic}
\end{algorithm}

\begin{algorithm}[htbp]
\caption{Calculate Pressure Index at any stage during a run chase}
\label{alg:pi-calculate-bbb}
\begin{algorithmic}[1]
\Require Target $T$
\Require Total balls $B$
\Require Runs scored $R'$
\Require Balls faced $B'$
\Require Wickets fallen $w_1, w_2, \ldots, w_k$
\Require Resources used $RU$
\Ensure Pressure Index $PI$

\State Calculate Initial Required Run Rate: $IRRR \leftarrow \frac{T \times 6}{B}$
\State Calculate Current Required Run Rate: $CRRR \leftarrow \frac{(T - R') \times 6}{B - B'}$
\State Initialize wicket weight sum: $\sum w_i \leftarrow 0$
\For{each fallen wicket at position $i$}
    \State $\sum w_i \leftarrow \sum w_i + w_i$ \Comment{Use weights from Table~\ref{tab:lemmer}}
\EndFor
\State Calculate: $PI \leftarrow \frac{CRRR}{IRRR} \times \frac{1}{2}\left[e^{\frac{RU}{100}} + e^{\frac{\sum w_i}{11}}\right]$

\Return $PI$
\end{algorithmic}
\end{algorithm}

\begin{algorithm}[htbp]
\caption{Generate over-by-over strategic recommendations}
\label{alg:recommendations}
\begin{algorithmic}[1]
\Require Current over $t$
\Require PI history $(PI_1, \ldots, PI_t)$
\Require Target $T$
\Require Wickets lost $w$
\Require Home/Away status
\Require Trained phase-wise models $\{\mathcal{M}_{\text{PP}}, \mathcal{M}_{\text{MO}}, \mathcal{M}_{\text{DO}}\}$
\Ensure Strategic recommendation and target PI range

\State Determine current phase $\phi$ based on over $t$
\State Select appropriate model $\mathcal{M}_{\phi}$ and thresholds from Table~\ref{tab:pi-thresholds-recommendations}
\State Predict next PI: $\hat{PI}_{t+1} \leftarrow \mathcal{M}_{\phi}(PI_{t-2}, PI_{t-1}, PI_t)$

\If{$\hat{PI}_{t+1} \in [0, 0.5)$}
    \State Recommendation $\leftarrow$ "Target Zone - Maintain aggressive scoring"
    
\ElsIf{$\hat{PI}_{t+1} \in [0.5, 1.5)$}
    \State Recommendation $\leftarrow$ "Acceptable Zone - Continue current approach"
    
\ElsIf{$\hat{PI}_{t+1} \in [1.5, 2.5)$ AND phase = Death Overs}
    \State Recommendation $\leftarrow$ "Acceptable/Risky - Accelerate carefully"
\Else
    \State Recommendation $\leftarrow$ "Avoid Zone - High risk, need immediate acceleration"
    \State Calculate required run rate adjustment
\EndIf

\State Display: Current PI, Predicted PI range, Recommendation, Required runs per over

\Return Recommendation with actionable targets
\end{algorithmic}
\end{algorithm}

\end{document}